\documentclass[journal,comsoc]{IEEEtran}
\usepackage{ifpdf}
\usepackage{cite}
\usepackage[cmex10]{amsmath}
\usepackage{array}
\usepackage{amsfonts}
\usepackage{mathrsfs}
\usepackage{arydshln}
\usepackage{graphicx}
\usepackage{float}
\usepackage{subfigure}
\usepackage{amssymb}
\usepackage{amsmath}
\usepackage{multirow}
\usepackage{multicol}
\usepackage{cite}
\usepackage[subfigure]{graphfig}
\usepackage{xcolor}
\usepackage{setspace}
\usepackage[commentsnumbered,boxed]{algorithm2e}
\usepackage{paralist}
\usepackage{enumitem}

\newtheorem{proposition}{\underline{Proposition}}

\newcommand{\qed}{\nobreak \ifvmode \relax \else
      \ifdim\lastskip<1.5em \hskip-\lastskip
      \hskip1.5em plus0em minus0.5em \fi \nobreak
      \vrule height0.75em width0.5em depth0.25em\fi}

\begin{document}
\title{{Constant Envelope Precoding with Adaptive Receiver Constellation in MISO Fading Channel}
\thanks{This work has been presented in part at the IEEE Global Communications Conference (GLOBECOM), San Diego, CA, USA, Dec. 6-10, 2015.}\thanks{S. Zhang is with the NUS Graduate School for Integrative Sciences and Engineering (NGS), National University of Singapore (e-mail:shuowen.zhang@u.nus.edu). She is also with the Department of Electrical and Computer Engineering, National University of Singapore.}\thanks{R. Zhang is with the
Department of Electrical and Computer Engineering, National
University of Singapore (e-mail:elezhang@nus.edu.sg). He is also
with the Institute for Infocomm Research, A*STAR, Singapore.}\thanks{T. J. Lim is with the Department of Electrical
and Computer Engineering, National University of Singapore
(e-mail:eleltj@nus.edu.sg).}}

\author{\IEEEauthorblockN{Shuowen~Zhang,~\IEEEmembership{Student Member,~IEEE}, Rui~Zhang,~\IEEEmembership{Senior Member,~IEEE}, and Teng Joon~Lim},~\IEEEmembership{Senior Member,~IEEE}}

\maketitle
\begin{abstract}
Constant envelope (CE) precoding is an appealing transmission technique which enables the realization of high power amplifier (PA) efficiency. For CE precoding in a single-user multiple-input single-output (MISO) channel, a desired constellation is feasible at the receiver if and only if it can be scaled to lie in an annulus, whose boundaries are characterized by the instantaneous channel realization. Therefore, if a fixed receiver constellation is used for CE precoding in a fading channel, where the annulus is time-varying, there is in general a non-zero probability of encountering a channel that makes CE precoding infeasible, thereby causing a high probability of error. To tackle this problem, this paper studies the adaptive receiver constellation design for CE precoding in a single-user MISO flat-fading channel with an arbitrary number of antennas at the transmitter. We first investigate the fixed-rate adaptive receiver constellation design to minimize the symbol error rate (SER). Specifically, an efficient algorithm is proposed to find the optimal amplitude-and-phase shift keying (APSK) constellation with two rings that is both feasible and of the maximum minimum Euclidean distance (MED), for any given constellation size and instantaneous channel realization. Numerical results show that by using the optimized fixed-rate adaptive receiver constellation, our proposed scheme achieves significantly improved SER performance over CE precoding with a fixed receiver constellation. Furthermore, based on the family of optimal fixed-rate adaptive two-ring APSK constellation sets, a variable-rate CE transmission scheme is proposed and numerically examined.
\end{abstract}
\begin{IEEEkeywords}
Constant envelope (CE) precoding, adaptive receiver constellation, amplitude-and-phase shift keying (APSK), minimum Euclidean distance (MED).
\end{IEEEkeywords}
\IEEEpeerreviewmaketitle
\section{Introduction}
Improving the power efficiency of radio frequency (RF) power amplifiers (PAs) reduces the energy consumption of wireless communication systems. From the perspective of power efficiency maximization, the most favorable input signals for PAs are \emph{constant envelope (CE)} signals. Specifically, since CE input signals have the lowest possible peak-to-average power ratio (PAPR), the backoff required for PA operation is minimized, and hence the power efficiency is maximized \cite{RFPA}.
It is also worth noting that CE input signals impose lower requirement on the dynamic range of the PAs than their non-CE counterparts, thereby requiring less expensive PAs to be used in practice.

To realize CE input signals, the complex baseband signal at each transmit antenna is required to have constant amplitude, and information can only be modulated in the transmitted signal phase. This stringent constraint, however, induces new signal processing challenges. For single-antenna channels, various CE modulation techniques have been studied, e.g., continuous phase modulation (CPM) \cite{digicom}, CE-OFDM (orthogonal frequency division multiplexing) \cite{CEOFDM}, etc. On the other hand, for multi-antenna channels, \emph{CE precoding} was recently proposed and investigated in  \cite{SUCE,JSTSPCE,MUCE,improved,freqCE,phaseconstraint}. Specifically, for the case of single-user multiple-input single-output (MISO) channels, it has been shown in \cite{SUCE,JSTSPCE} that by varying the signal phases at different transmit antennas, the noise-free signal at the single-antenna receiver always lies in an annular region (between two concentric rings), whose boundaries are determined by the instantaneous channel realization and per-antenna transmit power. In addition, low-complexity algorithms were proposed in \cite{SUCE,JSTSPCE} for CE precoding to achieve a nonlinear mapping from any desired received signal point within the annulus to the corresponding transmitted signal phases based on the instantaneous channel state information (CSI).\footnote{It is worth noting that there is another line of research on prototyping and performance analysis of the so-called outphasing or LINC (linear amplification with nonlinear components) technique (see, e.g., \cite{linearization,linc}), which can be viewed as a special case of CE precoding with two transmit antennas and one receive antenna.} Moreover, for multi-user large-scale MISO downlink systems \cite{overviewmassiveMIMO}, efficient CE precoding algorithms were developed in \cite{MUCE,improved} for frequency-flat channels and were shown to guarantee an arbitrarily low multi-user interference (MUI) power at each user receiver with a sufficiently large number of transmit antennas. The work in \cite{MUCE,improved} has been extended to the case with frequency-selective channels in \cite{freqCE}, where the transmitted signal phases for consecutive channel uses are jointly designed. Furthermore, \cite{phaseconstraint} extended the CE precoding scheme proposed in \cite{freqCE} to the case with an additional restriction on the phase difference between consecutive channel uses, in order to prevent the potential spectral regrowth caused by abrupt phase changes. Although the per-antenna CE constraint is clearly more restrictive than conventional average-based sum power constraint (SPC) and per-antenna power constraint (PAPC) (see, e.g., \cite{linearprecoding,dualityPAPC,multicellBD,LTCC,massivePAPC}), it was shown that with $M$ transmit antennas, an array power gain of $\mathcal{O}(M)$ is still achievable with the CE precoding schemes proposed in \cite{SUCE,JSTSPCE,MUCE,improved,freqCE,phaseconstraint}.

However, note that even for the simplest case of the single-user MISO channel, a desired constellation at the receiver is feasible with CE precoding if and only if it can be scaled to lie in the annular region, i.e., all the signal points in the scaled constellation can be mapped back to CE signals at the transmitter. Therefore, for a fading channel where the annulus is time-varying, a fixed receiver constellation may not always be feasible.\footnote{However, there are certain cases in which the CE precoding is always feasible regardless of channel fading. For example, PSK (phase shift keying) constellation is always feasible provided that the outer radius of the annulus is not zero. As another example, for a large-scale MISO channel with independent and identically distributed (i.i.d.) Rayleigh fading, the annulus was shown to become a disk region \cite{SUCE,JSTSPCE}, therefore any receiver constellation is feasible.} This can lead to severe performance degradation, which thus motivates this work on designing \emph{adaptive receiver constellations} based on the instantaneous CSI. There are generally two approaches for adaptive receiver constellation design \cite{wireless}, depending on whether the transmission rate or constellation size $N$ is fixed or adjustable for the given application:
\begin{itemize}[leftmargin=*]
\item For delay-constrained systems that require fixed-rate transmission (e.g., real-time voice and video), $N$ is fixed and the constellation $\mathcal{S}$ is adapted to channel conditions to minimize the receiver symbol error rate (SER);
\item For delay-tolerant systems that allow for variable-rate transmission, both $N$ and $\mathcal{S}$ can be jointly adapted based on the CSI to maximize the average transmission rate subject to a given SER requirement at the receiver.
\end{itemize}
To the best of our knowledge, neither approach in the above has been addressed in the literature for CE precoding.

In this paper, we study the adaptive receiver constellation design for CE precoding in a single-user MISO flat-fading channel with arbitrary number of antennas at the transmitter. Both cases of fixed-rate and variable-rate transmissions are considered. Our main contributions are summarized as follows:
\begin{itemize}[leftmargin=*]
\item First, we derive the \emph{fixed-rate} adaptive receiver constellation design that minimizes the SER at the receiver. This problem belongs to the class of circle packing problems that are known to be NP-hard \cite{circlepacking}. Hence, we approximate the exact SER by its union bound and assume that the constellation takes the form of amplitude-and-phase shift keying (APSK) so as to find a tractable solution. An efficient algorithm is proposed to obtain the optimal APSK constellation with two rings that is both feasible and maximizes the minimum Euclidean distance (MED) for SER minimization, given any constellation size and instantaneous CSI. Note that unlike existing works on APSK constellation design (see, e.g., \cite{digitalAPSK,turbocoded}) which consider an average power constraint of the signal points, our design is subjected to peak and minimum power constraints to ensure the constellation feasibility. It is also worth noting that for a given constellation size, our proposed design depends only on the ratio of the inner and outer radii of the annulus resulting from the instantaneous channel realization, and yields only a finite number of constellations, each corresponding to a certain range of this ratio. Therefore, the adaptive constellation set for any desired constellation size can be designed offline and stored at the transmitter/receiver for real-time transmission. To further reduce the amount of memory required for storage, which increases with the number of constellations resulting from optimization, we also propose a suboptimal design of the adaptive two-ring APSK constellation set which consists of fewer constellations than the optimal set.
\item Next, for the case of fixed-rate transmission, we present numerical results which show that our proposed scheme significantly outperforms CE precoding with fixed receiver constellation in terms of average SER, which is due to our constellation adaptation and optimization tailored for CE precoding.\footnote{For example, the proposed optimal $16$-APSK with two rings has an MED gain over rectangular $16$-QAM ($0.5411$ versus $0.4714$) when the latter is feasible.} It is also shown by numerical results that for our proposed scheme, replacing the optimal constellation set with the suboptimal one only leads to small performance loss that is almost negligible. Moreover, the effect of imperfect CSI at the transmitter (CSIT) on the performance of our proposed schemes is also investigated.
\item Finally, based on the family of optimal fixed-rate adaptive two-ring APSK constellation sets, we propose a \emph{variable-rate} transmission scheme for CE precoding. According to the instantaneous CSI, the constellation size is selected as the maximum value from its feasible set that yields the SER union bound lower than a target value, to maximize the transmission rate. The performance of the proposed scheme is numerically examined in terms of average spectral efficiency and compared with that of another variable-rate CE transmission scheme based on the family of rectangular QAM (quadrature amplitude modulation) constellations.
\end{itemize}

The remainder of this paper is organized as follows. Section \ref{secsys} presents the system model for CE precoding. Section \ref{secpf} presents the problem formulation of fixed-rate adaptive receiver constellation design, then Section \ref{secopt} and Section V present the optimal solution and a suboptimal solution, respectively. Numerical results are shown in Section \ref{secnum}, where the variable-rate transmission scheme is also presented. Finally, Section \ref{seccon} concludes the paper.

\textit{Notations}: Scalars and vectors are denoted by lower-case letters and boldface lower-case letters, respectively. ${\bf{z}}^T$ denotes the transpose of a vector ${\bf{z}}$. $\|{\bf{z}}\|_1$ and $\|{\bf{z}}\|_\infty$ are the $l_1$-norm and $l_\infty$-norm of a vector ${\bf{z}}$, respectively. $|z|$ is the absolute value of a scalar $z$. $\mathbb{C}^{m\times n}$ denotes the space of $m\times n$ complex matrices. The distribution of a circularly symmetric complex Gaussian (CSCG) random variable with mean $\mu$ and variance $\sigma^2$ is denoted by $\mathcal{CN}(\mu,\sigma^2)$; and $\sim$ stands for ``distributed as''. $\mathrm{Prob}(\cdot)$ denotes the probability. $\max\{x,y\}$ denotes the maximum between two real numbers $x$ and $y$. $|\mathcal{X}|$ denotes the cardinality of a set $\mathcal{X}$. $\mathcal{X}\cup\mathcal{Y}$ denotes the union of two sets $\mathcal{X}$ and $\mathcal{Y}$.
\section{System Model}\label{secsys}
We consider a single-user MISO flat-fading channel with $M$ antennas at the transmitter and one antenna at the receiver. The baseband received signal is given by
\begin{equation}
y={\bf{h}}^T{\bf{x}}+n,\label{channel}
\end{equation}
where ${\bf{x}}$ denotes the $M\times1$ transmitted signal vector; ${\bf{h}}\in \mathbb{C}^{M\times 1}$ denotes the channel vector, which is assumed to be perfectly known at both the transmitter and receiver unless specified otherwise; $n\sim \mathcal{CN}(0,\sigma^2)$ denotes the CSCG noise at the receiver. Assume a total transmit power denoted by $P$, which is equally allocated to the $M$ antennas. Under the per-antenna CE constraint, the transmitted signal at each antenna is expressed as
\begin{equation}
x_i=\sqrt{\frac{P}{M}}e^{j\theta_i},\quad i=1,...,M,\label{CEconstraint}
\end{equation}
where information is modulated in the transmitted signal phases $\theta_i\in[0,2\pi),i=1,...,M$. By varying $\theta_i$'s, the noise-free received signal $d\overset{\Delta}{=}{\bf{h}}^T{\bf{x}}=\sqrt{\frac{P}{M}}\sum_{i=1}^M
h_ie^{j\theta_i}$ is shown in \cite{SUCE} and \cite{JSTSPCE} to always lie in the following region:
\begin{equation}
\mathcal{D}=\{d\in \mathbb{C}:r\leq |d|\leq R\},\label{region}
\end{equation}
where $R=\sqrt{\frac{P}{M}}\|{\bf{h}}\|_1$ and $r=\sqrt{\frac{P}{M}}\max\{2\|{\bf{h}}\|_\infty-\|{\bf{h}}\|_1,0\}$.

\begin{figure}[htb]
  \centering
  \includegraphics[width=7.5cm]{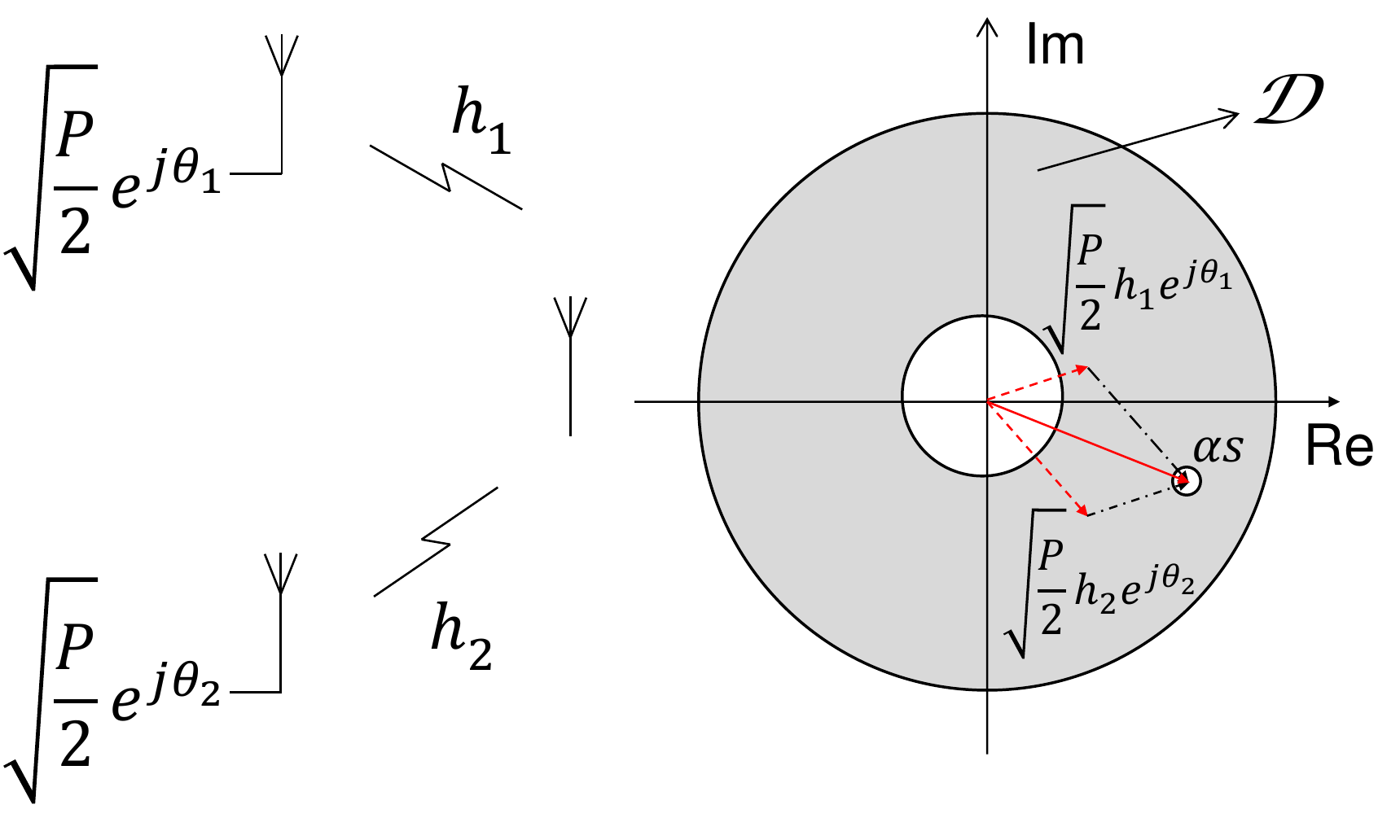}
  \caption{System diagram for CE precoding with $M=2$.}\label{CEdiagram}
\end{figure}

As a result, any desired received signal constellation $\mathcal{S}$ (e.g., QAM) is feasible if and only if there exists a scaling factor $\alpha>0$ such that $\alpha \mathcal{S}\subset \mathcal{D}$. In other words, $\mathcal{S}$ is feasible if and only if $\frac{r}{R}\leq {\underset{s\in \mathcal{S}}{\min} |s|}\slash{\underset{s\in \mathcal{S}}{\max} |s|}$. For any feasible $\mathcal{S}$, the transmitted phase set $\{\theta_i\}$ corresponding to any signal point $s\in \mathcal{S}$ at the receiver and any $\alpha$ such that $d=\alpha s$ can be readily obtained by the algorithms proposed in \cite{SUCE} and \cite{JSTSPCE}, which are thus omitted for brevity. The channel in (\ref{channel}) is thus equivalently represented by
\begin{equation}
y=\alpha s+n.\label{equichannel}
\end{equation}
For illustration, Fig. \ref{CEdiagram} shows the system diagram for CE precoding with $M=2$ transmit antennas. Note that in order to maximize the received signal power and yet meet the feasibility condition, we should set $\alpha=R$ for any feasible constellation $\mathcal{S}$ with $\underset{s\in\mathcal{S}}{\max} |s|=1$, such that the signal point with the largest amplitude in $\mathcal{S}$ lies on the outer boundary of $\mathcal{D}$ at the receiver.

\begin{figure}[h]
  \centering
  \subfigure[Feasible case with $\frac{r}{R}\leq\frac{1}{3}$]{
    \includegraphics[width=3.5cm]{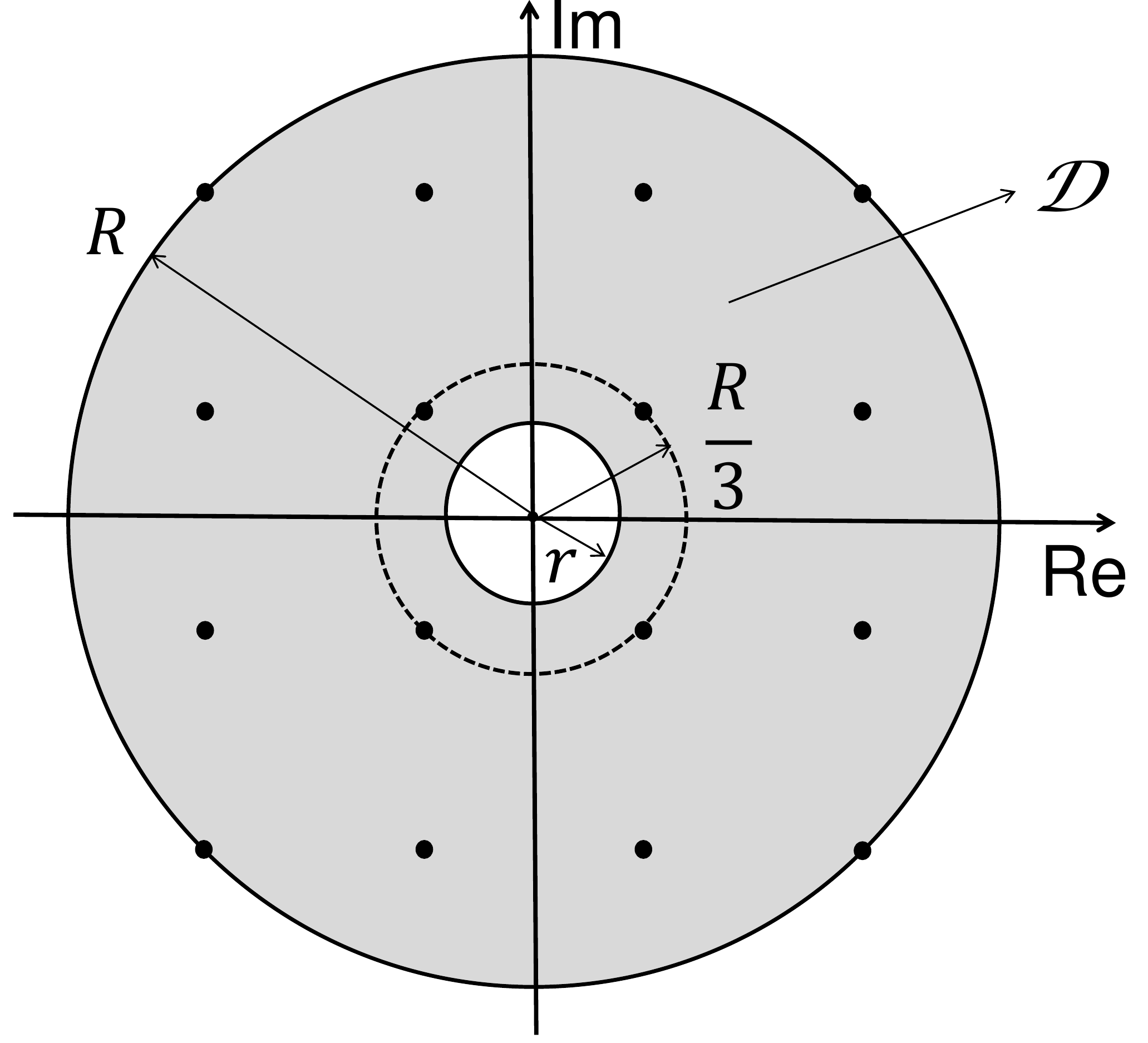}}
  \hspace{0in}
  \subfigure[Infeasible case with $\frac{r}{R}>\frac{1}{3}$]{
    \includegraphics[width=3.5cm]{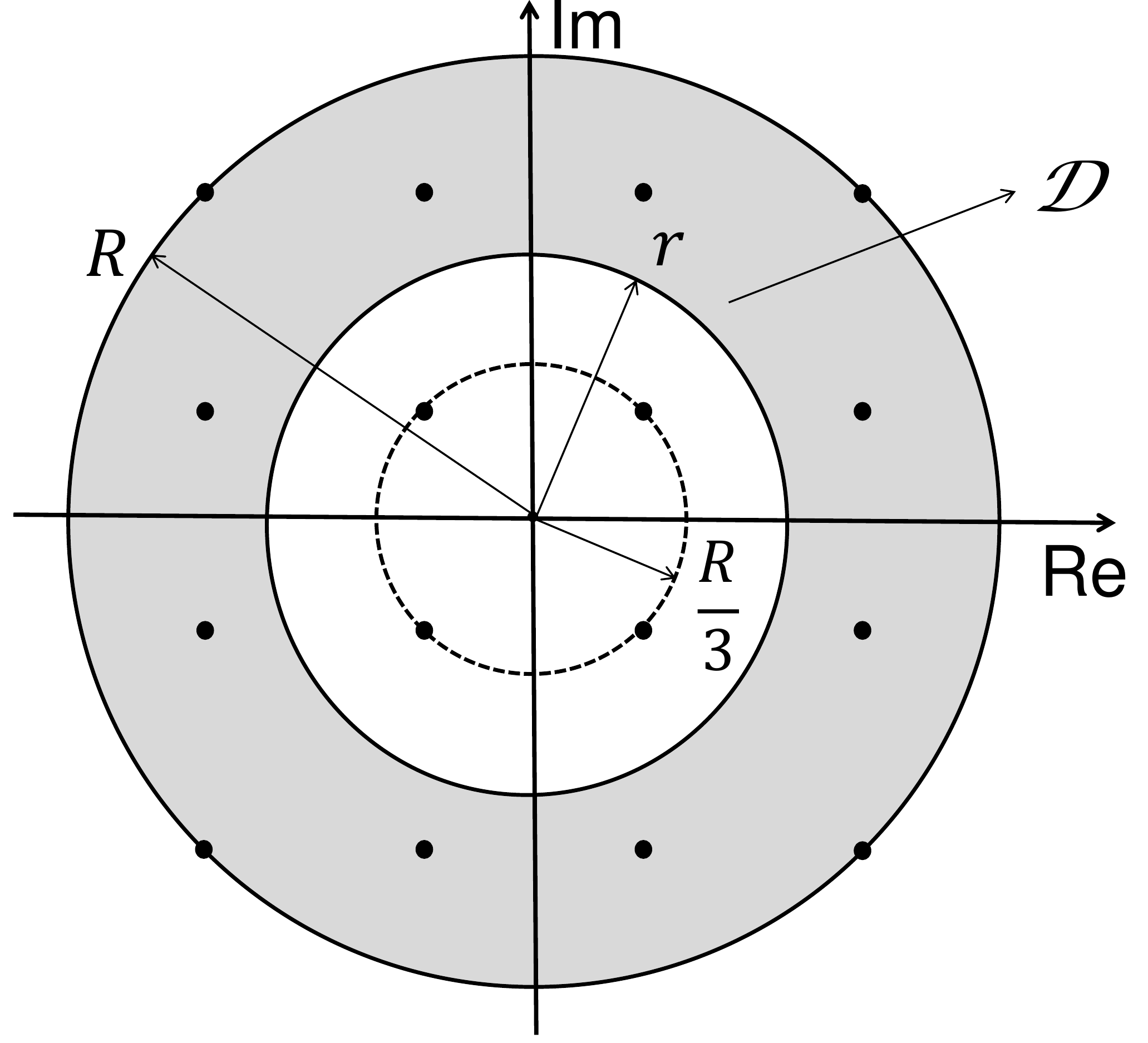}}
  \caption{Feasibility of rectangular 16-QAM with different channel realizations.}\label{feasibility}
\end{figure}

For MISO fading channel with time-varying $h_i$'s and as a result both time-varying $r$ and $R$, a fixed receiver constellation $\mathcal{S}$ may not always be feasible. For example, if $\mathcal{S}$ is a rectangular 16-QAM constellation (i.e., $\underset{s\in \mathcal{S}}{\min} |s|/{\underset{s\in \mathcal{S}}{\max} |s|}=\frac{1}{3}$), it is feasible only when $\frac{r}{R}\leq\frac{1}{3}$ holds; otherwise, it is infeasible, as shown in Fig. \ref{feasibility}. In particular, for the case where $M=2$ and $h_i$'s are i.i.d. zero-mean CSCG random variables (i.e., i.i.d. Rayleigh fading channels), it can be shown that $\mathrm{Prob}(\frac{r}{R}>\frac{1}{3})=0.4$,\footnote{In this case, the cumulative distribution function (CDF) of $\frac{r}{R}$ can be shown to be given by $\mathrm{Prob}(\frac{r}{R}\leq x)=\frac{2x}{1+x^2},\quad x\in[0,1]$.} i.e., rectangular 16-QAM constellation is infeasible with CE precoding for $40\%$ of channel realizations. This will result in severe performance degradation, which thus motivates our design of adaptive receiver constellation for CE transmission based on the instantaneous CSI.
\section{Problem Formulation}\label{secpf}
In this section, we formulate the problem of fixed-rate adaptive receiver constellation design for CE precoding. Our aim is to optimize the $N$-ary constellation $\mathcal{S}$ with fixed $N$ based on the instantaneous CSI to minimize the receiver SER. Note that for given $N$, our constellation design depends only on the ratio $\frac{r}{R}$ resulting from $h_i$'s, but not on the exact values of $h_i$'s or the value of $M$. Therefore, our proposed constellation design can be implemented offline and then used for real-time transmission. Also note that as will be shown later in the paper, there is only a finite set of constellations in our design for any given $N$, each corresponding to a certain range of the value $\frac{r}{R}$. Hence, the amount of memory required for storing the designed constellations at the transmitter/receiver is manageable.

However, this constellation optimization problem is in general challenging, since it belongs to the class of circle packing problems that are known to be NP-hard \cite{circlepacking}. Therefore, we make the following simplifications to obtain a tractable solution:
\begin{enumerate}[leftmargin=*]
\item Instead of considering the exact SER, $P_s$, at the receiver, we minimize the union bound of SER given by $P_s\leq (N-1)Q\left(\sqrt{\frac{(Rd_{\mathrm{min}})^2}{2\sigma^2}}\right)$ \cite{digicom}, where we assume that $\mathcal{S}$ is an equiprobable signal set with $\underset{s\in \mathcal{S}}{\max}|s|=1$, and the maximum likelihood (ML) detection is used at the receiver to recover the signal point in $\mathcal{S}$. This can be achieved by maximizing $d_{\mathrm{min}}$, which denotes the MED between any two constellation points in $\mathcal{S}$;
\item We assume that the constellation $\mathcal{S}$ follows an APSK structure, where signal points are distributed on concentric rings. APSK constellations are favorable due to the following reasons: first, they can be easily fitted into any annular region at the receiver by adjusting the radii of their rings; moreover, since only a limited number of parameters suffice to fully characterize an APSK constellation, adaptive APSK can be designed and implemented with low complexity, as will be shown later in this paper.
\end{enumerate}

Consider an $N$-ary APSK constellation $\mathcal{S}$ modeled by
\begin{align}
\mathcal{S}=\bigg\{\rho_le^{j\left(\frac{2\pi k_l}{N_l}+\omega_l\right)},\ k_l=0,...,N_l-1,\ l=1,...,L\bigg\},\label{APSKsignalset}
\end{align}
where $L\geq 1$ denotes the number of concentric rings; $\rho_l$ denotes the radius of the $l$th ring; $N_l\geq 1$ denotes the number of signal points that are uniformly spaced on the $l$th ring, with $\sum_{l=1}^L N_l=N$; $\omega_l$ denotes the phase offset of the $l$th ring with respect to the reference phase $\omega_1=0$. We assume the $L$ rings are indexed such that $\rho_L<\rho_{L-1}<\cdots<\rho_1$. Thus, we have $\rho_1=1$ since $\underset{s\in \mathcal{S}}{\max} |s|=1$ by assumption.

Our objective is to maximize the MED of $\mathcal{S}$ by jointly optimizing $L$, $\{N_l\}_{l=2}^L$, $\{\rho_l\}_{l=2}^L$ and $\{\omega_l\}_{l=2}^L$, subject to all the constellation points lying in an annulus with inner and outer boundary radii given by $\frac{r}{R}$ and $1$, respectively. Let $\mathcal{L}=\{1,2,...,L\}$ denote the set of rings. We formulate the following optimization problem with given $N$ and $\frac{r}{R}$ as
\begin{align}
\mbox{(P1)} \underset{\scriptstyle L, \{N_l\}_{l=2}^L, \atop \scriptstyle \{\rho_l\}_{l=2}^L,\{\omega_l\}_{l=2}^L,d_{\mathrm{min}}}{\max} &d_{\mathrm{min}} \nonumber\\
\text{s.t.}\quad  &d_{\mathrm{min}}\leq d_l,\ \forall l\in \mathcal{L} \label{dl}\\
&d_{\mathrm{min}}\leq d_{l,h},\ \forall l,h\in \mathcal{L},\ l\neq h\\
&\frac{r}{R}\leq \rho_L<\rho_{L-1}<\cdots<\rho_2< 1\label{powerc}\\
&\sum_{l=1}^L N_l=N\\
&N_l\in \mathbb{Z}^+,\ \forall l\in \mathcal{L}\\
&\omega_l\in[0,2\pi),\ \forall l\in \mathcal{L}\backslash \{1\}\\
&L\in \mathbb{Z}^+,\label{P1}
\end{align}
where $\mathbb{Z}^+$ is the set of positive integers; $d_l$ denotes the intra-ring MED between any two signal points on the $l$th ring; $d_{l,h}$ denotes the inter-ring MED between any two signal points that are located on the $l$th and $h$th rings, respectively. From (\ref{APSKsignalset}), $d_l$ can be obtained as
\begin{equation}
d_l=
\begin{cases}
\rho_l\sqrt{2\left(1-\cos\left(\frac{2\pi}{N_l}\right)\right)}\quad &N_l\geq2\\
\infty\quad &N_l=1.
\end{cases}\label{in}
\end{equation}
Note that when $N_l=1$, $d_l=\infty$ and the corresponding constraint in (\ref{dl}) of (P1) can be removed for ring $l$. Using the cosine rule, we can also obtain
\begin{equation}
\begin{aligned}
d_{l,h}=\sqrt{\rho_l^2+\rho_h^2-2\rho_l\rho_hC_{l,h}(N_l,N_h,\omega_l,\omega_h)}, \ l\neq h,\label{adjacent}
\end{aligned}
\end{equation}
where
\begin{align}
C_{l,h}&(N_l,N_h,\omega_l,\omega_h)=\underset{m,n}{\max} \cos\left(\frac{2\pi n}{N_{l}}+\omega_{l}-\omega_{h}-\frac{2\pi m}{N_h}\right)\nonumber\\
&m\in\{0,1,...,N_h-1\},\quad n\in\{0,1,...,N_l-1\}.\label{C}
\end{align}

In this paper, we solve (P1) for the special case of $L=2$,\footnote{Note that for the case of $L=1$, the solution of (P1) can be easily shown to be the conventional $N$-ary PSK. For consistency, we also consider this case as a two-ring APSK constellation.} while our proposed method can also be extended to the more general case of $L>2$ in principle. With $L=2$, Problem (P1) is simplified as
\begin{align}
\mbox{(P2)}\underset{\rho_2,\omega_2,N_2,d_{\mathrm{min}}}{\max}\ &d_{\mathrm{min}} \nonumber\\
\text{s.t.}\quad  &d_{\mathrm{min}}-\sqrt{2B_1}\leq0\\
&d_{\mathrm{min}}-\rho_2\sqrt{2B_2}\leq0\\
&d_{\mathrm{min}}-\sqrt{1+\rho_2^2-2\rho_2 C_{1,2}(N_2,\omega_2)}\leq0 \label{3cP2}\\
&\frac{r}{R}\leq \rho_2\leq 1\\
&\omega_2\in[0,2\pi)\\
&N_2\in\{1,...,N-1\},\label{P2}
\end{align}
with $N_1=N-N_2$, $B_l=\begin{cases}1-\cos\left(\frac{2\pi}{N_l}\right), N_l\geq2\\ \infty, N_l=1\end{cases}$ for $l\in\{1,2\}$, and $C_{1,2}(N_2,\omega_2)=\underset{m,n}{\max} \cos\left(\frac{2\pi n}{N_{2}}+\omega_{2}-\frac{2\pi m}{N_1}\right),$ $m\in\{0,1,...,N_1-1\},\ n\in\{0,1,...,N_2-1\}$.
\section{Optimal Solution to Problem (P2)}\label{secopt}
Note that Problem (P2) is a non-convex optimization problem since the constraint in (\ref{3cP2}) is non-convex due to the coupled $\rho_2$ and $\omega_2$, and the fact that $N_2$ is an integer variable. In this section, we propose an efficient algorithm for solving Problem (P2). Specifically, we first optimize $\omega_2$ and $\rho_2$ for each given $N_2\in\{1,2,...,N-1\}$, and denote the corresponding optimal value of Problem (P2) by $d_{\mathrm{min}}^*(N_2)$. Then, we find the optimal $N_2$ as $N_2^*=\underset{N_2\in\{1,2,...,N-1\}}{\arg\max} d_{\mathrm{min}}^*(N_2)$ via one-dimensional search over $N_2$.

First, we show one property of $N_2^*$ in the following proposition, which helps reduce its search space.
\begin{proposition}\label{propN2}
For Problem (P2), it holds that ${N_2^*}\leq\frac{N}{2}$,\footnote{In the sequel, we follow the convention to assume that $N$ is an integral power of two, thus is an even integer.}  i.e., there should be no more points allocated to the inner ring than to the outer ring.
\end{proposition}
\begin{IEEEproof}
Please refer to Appendix \ref{proofpropN2}.
\end{IEEEproof}

With Proposition \ref{propN2}, the set of potentially optimal values of $N_2$ is reduced to $\left\{1,2,...,\frac{N}{2}\right\}$, and the complexity of the one-dimensional search for $N_2^*$ is thus reduced by half.

Then, let $\omega_2^*(N_2)$ and $\rho_2^*(N_2)$ denote the solution to Problem (P2) with a given $N_2\in\left\{1,2,...,\frac{N}{2}\right\}$. Notice that with given $N_2$, $\omega_2^*(N_2)$ can be first obtained by solving the following problem:
\begin{equation}\nonumber
\begin{aligned}
(\mbox{P2.1})\quad \underset{0\leq\omega_2<2\pi}{\min} C_{1,2}(N_2,\omega_2).
\end{aligned}
\end{equation}
Problem (P2.1) can be optimally solved by Algorithm 1 shown in Table \ref{algo1} at the top of next page, for which the details are given in Appendix \ref{proofalgo1}.
\newpage

\begin{table}[h]
\caption{Algorithm 1: Algorithm for solving Problem (P2.1)}\label{algo1}
\end{table}
\vspace{-10mm}
\begin{algorithm}[htb]
\SetKwData{Index}{Index}
\KwIn{$N$, $N_2$}
\KwOut{$\omega_2^*(N_2)$, $C_{1,2}^*(N_2)$}

Set $N_1=N-N_2$. Initialize $\mathcal{X}=\left\{-\frac{2\pi}{N_1}\right\}$.

\For {$m=0$ \KwTo $N_1-1$}{
\For {$n=0$ \KwTo $N_2-1$}{
\If {$-\frac{2\pi}{N_1} <\frac{2\pi n}{N_2}-\frac{2\pi m}{N_1}\leq 0$}{
 $\mathcal{X}=\mathcal{X} \cup \left\{\frac{2\pi n}{N_2}-\frac{2\pi m}{N_1}\right\}$;}
}}
Set $K=|\mathcal{X}|$. Sort the elements in $\mathcal{X}$ in a descending order: $X_{(1)}> X_{(2)}>...> X_{(K)}$.

{Set $k^*=\underset{1\leq k\leq K-1}{\arg\min}\frac{X_{(k+1)}-X_{(k)}}{2}$; \\
$\quad\ \omega_2^*(N_2)=\frac{-X_{(k^*)}-X_{(k^*+1)}}{2}$;\\
$\quad\ C_{1,2}^*(N_2)=\cos\left(\frac{X_{(k^*)}-X_{(k^*+1)}}{2}\right)$.}
\end{algorithm}

Let $C_{1,2}^*(N_2)$ denote the optimal value of Problem (P2.1). Note that in Table \ref{algo1}, since ${X_{(k)}-X_{(k+1)}}\in\left(0,\frac{2\pi}{N_1}\right]$ for $\forall k\in\{1,2,...,K-1\}$, we have $C_{1,2}^*(N_2)\in\left[\cos\left(\frac{\pi}{N_1}\right),1\right)$. Next, we obtain $d_{\mathrm{min}}^*(N_2)$ by finding $\rho_2^*(N_2)$ through solving the following problem:
\begin{equation}\nonumber
\begin{aligned}
&\mbox{(P2.2)}\\
&\underset{\frac{r}{R}\leq\rho_2\leq1}{\max} {\min} \left\{\sqrt{2B_1},\rho_2\sqrt{2B_2}, \sqrt{1+\rho_2^2-2\rho_2C_{1,2}^*(N_2)}    \right\}.
\end{aligned}
\end{equation}

Problem (P2.2) is still non-convex with respect to $\rho_2$. To find its optimal solution, we observe that the inter-ring MED $\sqrt{1+\rho_2^2-2\rho_2C_{1,2}^*(N_2)}$ first decreases with $\rho_2$ when $\rho_2< C_{1,2}^*(N_2)$, and then increases with $\rho_2$ when $C_{1,2}^*(N_2)\leq\rho_2\leq1$. Furthermore, the intra-ring MED of the inner ring, $\rho_2\sqrt{2B_2}$,  strictly increases with  $\rho_2$. Based on the above results, we discuss the solution of Problem (P2.2) in the following two cases.

\begin{itemize}[leftmargin=*]
\item Case 1: $C_{1,2}^*(N_2)\leq\frac{r}{R}$. In this case, we have $\rho_2\geq\frac{r}{R}\geq C_{1,2}^*(N_2)$, and the objective function of Problem (P2.2) is a non-decreasing function of $\rho_2$. Therefore, we have
\begin{align}
&\rho_2^*(N_2)=1,\nonumber\\
&d_{\mathrm{min}}^*(N_2)=\min\left\{\sqrt{2B_1},\sqrt{2B_2},\sqrt{2-2C_{1,2}^*(N_2)}\right\}.\label{case1}
\end{align}
This is consistent with our intuition that as $\frac{r}{R}$ becomes large, the two rings converge and allocating all the signal points on the outer ring is optimal.
\item Case 2: $C_{1,2}^*(N_2)> \frac{r}{R}$. In this case, we further divide the feasible range of $\rho_2$ into $\left[\frac{r}{R},C_{1,2}^*(N_2)\right]$ and $\left(C_{1,2}^*(N_2),1\right]$, which are referred to as region I and region II, respectively. We then investigate the locally optimal $\rho_2$ within the two regions, which are denoted by $\rho_{2,\mathrm{I}}^*(N_2)$ and $\rho_{2,\mathrm{II}}^*(N_2)$, respectively. Let $d_{\mathrm{min,I}}^*(N_2)$ and $d_{\mathrm{min,II}}^*(N_2)$ denote the local optimum of Problem (P2.2) for the two regions, respectively. The globally optimal solution to Problem (P2.2) is then given by
\begin{equation}
\rho_2^*(N_2)=\begin{cases} \rho_{2,\mathrm{I}}^*(N_2)\quad d_{\mathrm{min,I}}^*(N_2)\geq d_{\mathrm{min,II}}^*(N_2)\\
\rho_{2,\mathrm{II}}^*(N_2)\quad {\rm{otherwise}}.\label{case2}
\end{cases}
\end{equation}
Furthermore, for region I, since $\sqrt{1+\rho_2^2-2\rho_2C_{1,2}^*(N_2)}$ and $\rho_2\sqrt{2B_2}$ are monotonically decreasing and increasing with $\rho_2$, respectively, the shape of the objective function of Problem (P2.2) as well as the corresponding $\rho_{2,\mathrm{I}}^*(N_2)$ are dependent on the values of $B_1$, $B_2$, $C_{1,2}^*(N_2)$ and $\frac{r}{R}$. We thus propose Algorithm 2 in Table \ref{algo2} to find $\rho_{2,\mathrm{I}}^*(N_2)$, for which the details are given in Appendix \ref{appendixalgo2}.
\begin{table}[!htb]
\caption{Algorithm 2: Algorithm for finding $\rho_{2,\mathrm{I}}^*(N_2)$}\label{algo2}
\end{table}
\vspace{-30pt}
\begin{algorithm}[!htb]
\SetKwData{Index}{Index}
\KwIn{$B_1$, $B_2$, $C_{1,2}^*(N_2)$, $\frac{r}{R}$}
\KwOut{$\rho_{2,\mathrm{I}}^*(N_2)$, $d_{\mathrm{min,I}}^*(N_2)$}
Obtain $\bar{\rho}_2$ by (\ref{brho2}).

\eIf {$\bar{\rho}_2\in\left[\frac{r}{R},C_{1,2}^{*}(N_2)\right]$}
{\eIf {$\bar{\rho}_2\leq\sqrt{\frac{B_1}{B_2}}$}{$\rho_{2,\mathrm{I}}^*(N_2)=\bar{\rho}_2$;\quad $d_{\mathrm{min,I}}^*(N_2)=\sqrt{2B_2}\bar{\rho}_2$;\ (Case i)}{$\rho_{2,\mathrm{I}}^*(N_2)=C_{1,2}^*(N_2)-\sqrt{C_{1,2}^{*2}(N_2)-1+2B_1}$;
\quad $d_{\mathrm{min,I}}^*(N_2)=\sqrt{2B_1}$;\quad (Case ii)}}
{
\eIf {$\sqrt{2B_1}\geq \sqrt{\left(\frac{r}{R}\right)^2+1-2\left(\frac{r}{R}\right)C_{1,2}^*(N_2)}$}
{$\rho_{2,\mathrm{I}}^*(N_2)=\frac{r}{R}$;
\quad $d_{\mathrm{min,I}}^*(N_2)=\sqrt{\left(\frac{r}{R}\right)^2+1-2\left(\frac{r}{R}\right)C_{1,2}^*(N_2)}$; \quad(Case iii)}
{$\rho_{2,\mathrm{I}}^*(N_2)=C_{1,2}^*(N_2)-\sqrt{C_{1,2}^{*2}(N_2)-1+2B_1}$;\quad $d_{\mathrm{min,I}}^*(N_2)=\sqrt{2B_1}$.\quad (Case iv)}
}
\end{algorithm}

On the other hand, for region II, similar to Case 1 above, the objective function of Problem (P2.2) is a non-decreasing function of $\rho_2$ since $\rho_2>C_{1,2}^*(N_2)$. Therefore, we have
\begin{align}
&\rho_{2,\mathrm{II}}^*(N_2)=1,\nonumber\\ &d_{\mathrm{min,II}}^*(N_2)=\min\left\{\sqrt{2B_1},\sqrt{2B_2},\sqrt{2-2C_{1,2}^*(N_2)}\right\}.\label{II}
\end{align}
\end{itemize}

So far, we have found the optimal solution for $\omega_2^*(N_2)$, $\rho_2^*(N_2)$ and the corresponding maximum MED $d_{\mathrm{min}}^*(N_2)$ for any given $N_2$. With Proposition \ref{propN2}, the optimal number of points on the inner ring is then obtained as $N_2^*=\underset{N_2\in\left\{1,2,...,\frac{N}{2}\right\}}{\arg\max} d_{\mathrm{min}}^*(N_2)$. This completes our proposed algorithm for Problem (P2), which is summarized as Algorithm 3 shown in Table \ref{algo 3} at the top of next page. For any given $N$ and $\frac{r}{R}$, this algorithm can be shown to find the optimal solution to Problem (P2) with worst-case complexity of $\mathcal{O}(N^3\log N)$.
\vspace{-5mm}

\begin{table}[htb]
\caption{Algorithm 3: Algorithm for solving Problem (P2)}\label{algo 3}
\end{table}
\vspace{-25pt}
\begin{algorithm}[htb]
\SetKwData{Index}{Index}
\KwIn {$N$, $\frac{r}{R}$}
\KwOut {$N_2^*$, $\omega_2^*$, $\rho_2^*$}
\For{$N_2=1$ \KwTo $\frac{N}{2}$}{
Compute $B_1$ and $B_2$.

Obtain $\omega_2^*(N_2)$ and $C_{1,2}^*(N_2)$ by Algorithm 1.

\eIf {$C_{1,2}^*(N_2)\leq\frac{r}{R}$}
{ Obtain $\rho_2^*(N_2)$ and $d_{\mathrm{min}}^*(N_2)$ by (\ref{case1}).}
{
Obtain $\rho_{2,\mathrm{I}}^*(N_2)$ and $d_{\mathrm{min,I}}^*(N_2)$ by Algorithm 2.

Obtain $\rho_{2,\mathrm{II}}^*(N_2)$ and $d_{\mathrm{min,II}}^*(N_2)$ by (\ref{II}).

\eIf{$d_{\mathrm{min,I}}^*(N_2)\geq d_{\mathrm{min,II}}^*(N_2)$}
{$\rho_2^*(N_2)=\rho_{2,\mathrm{I}}^*(N_2)$; $d_{\mathrm{min}}^*(N_2)=d_{\mathrm{min,I}}^*(N_2)$.}
{$\rho_2^*(N_2)=\rho_{2,\mathrm{II}}^*(N_2)$; $d_{\mathrm{min}}^*(N_2)=d_{\mathrm{min,II}}^*(N_2)$.}
}
}

{Set $N_2^*=\underset{N_2\in\left\{1,2,...,\frac{N}{2}\right\}}{\arg\max} d_{\mathrm{min}}^*(N_2)$; $\rho_2^*=\rho_2^*(N_2^*)$; $\omega_2^*=\omega_2^*(N_2^*)$.}
\end{algorithm}

\begin{figure}[htb]
\centering
\includegraphics[width=7.5cm]{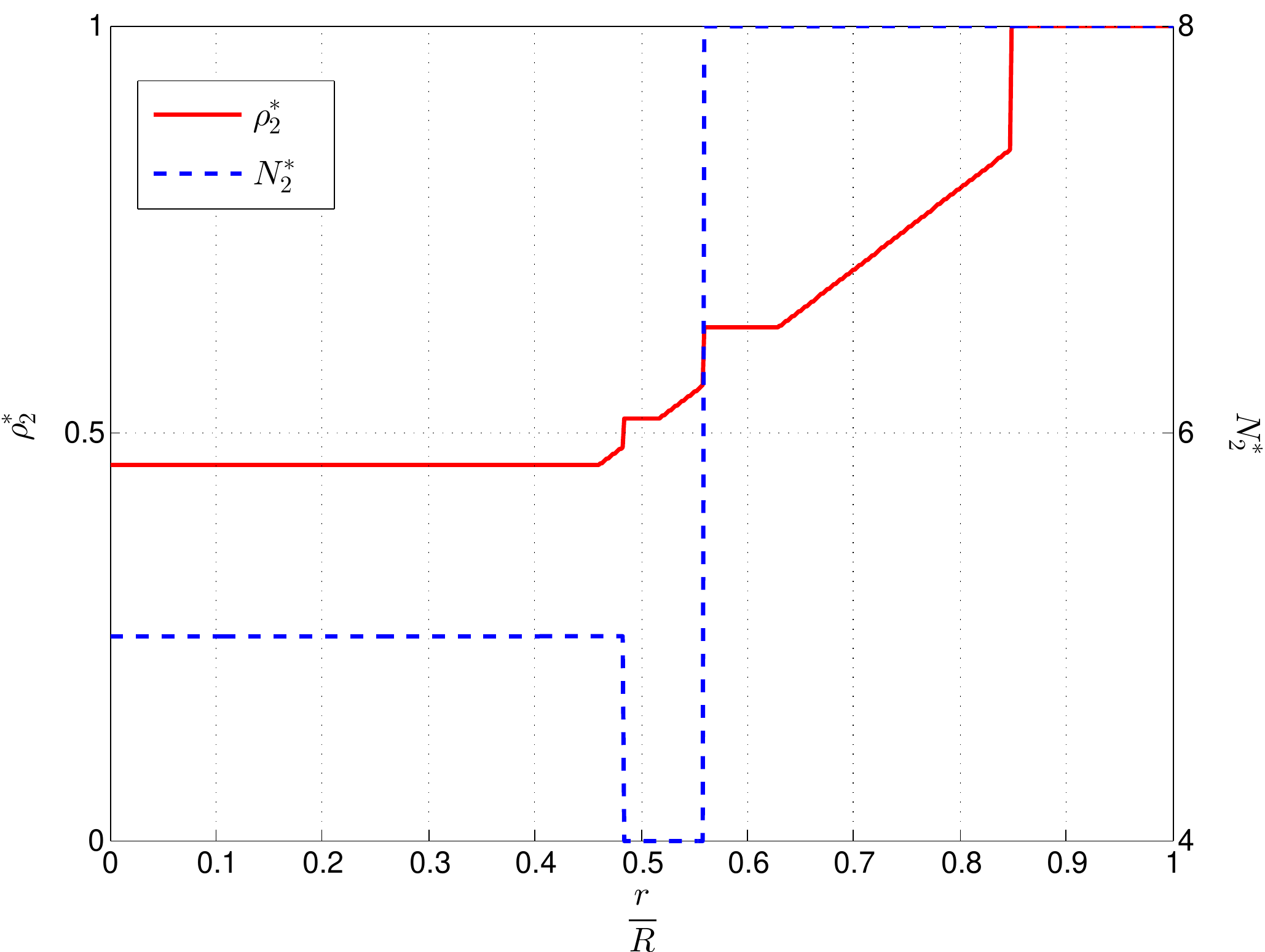}
\caption{$\rho_2^*$ and $N_2^*$ versus $\frac{r}{R}$ when $N=16$.}\label{rR}
\end{figure}

\begin{figure}[htb]
  \centering
  \subfigure[$\frac{r}{R}=0.4$]{
    \label{fig:subfig:a}
    \includegraphics[width=2.5cm]{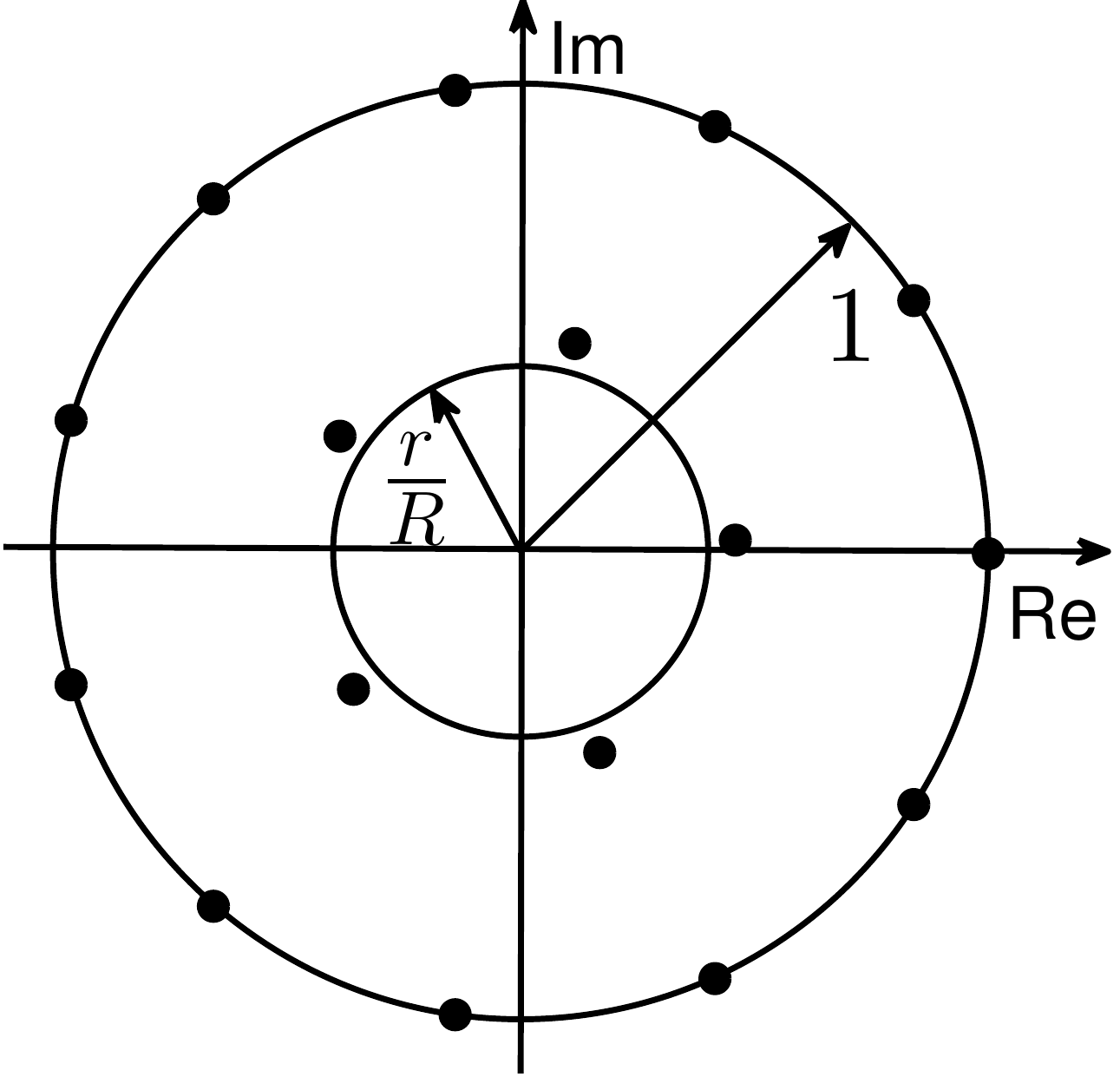}}
  \hspace{0in}
  \subfigure[$\frac{r}{R}=0.5$]{
    \label{fig:subfig:b}
    \includegraphics[width=2.5cm]{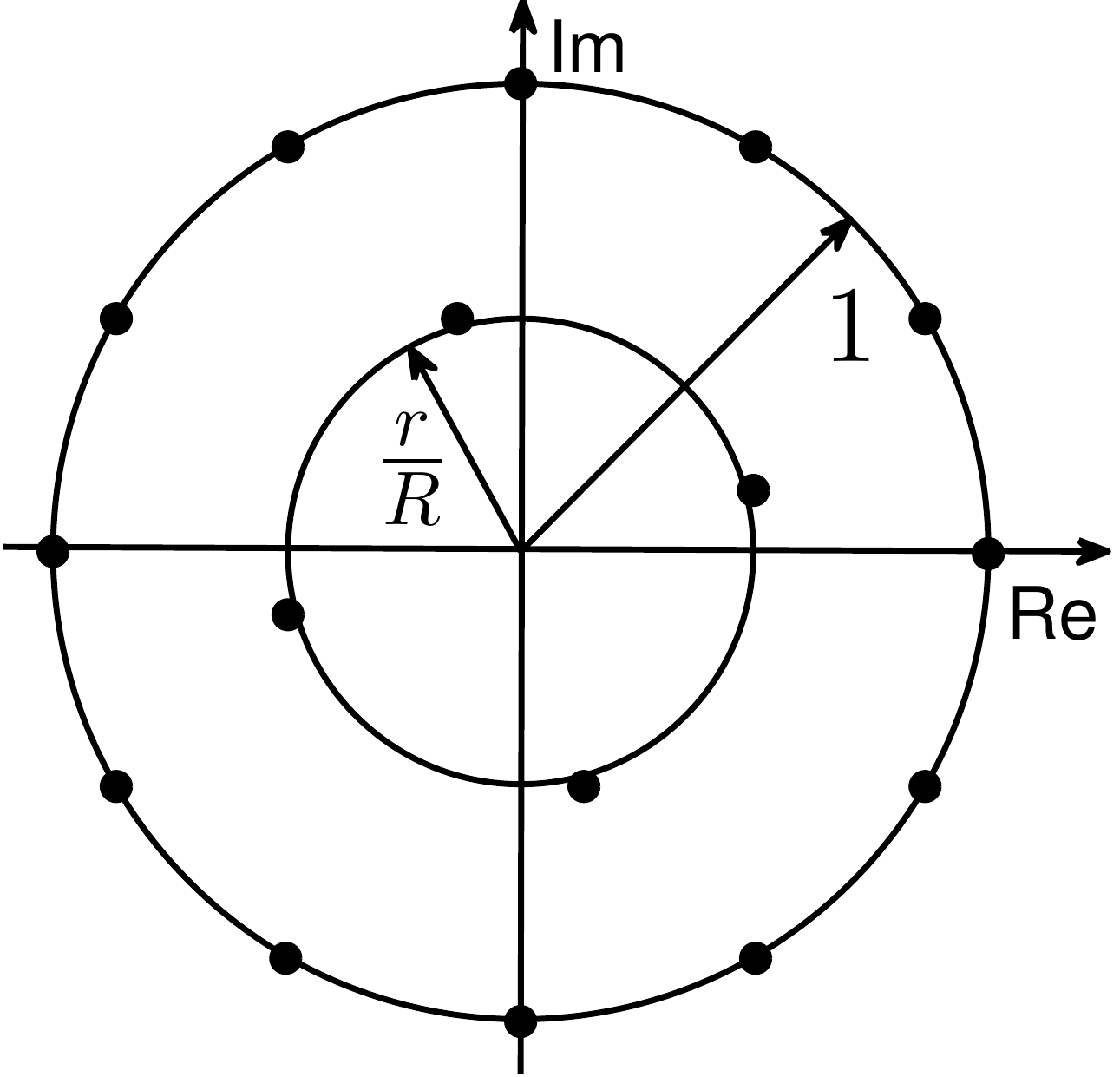}}
  \hspace{0in}
  \subfigure[$\frac{r}{R}=0.7$]{
    \label{fig:subfig:b}
    \includegraphics[width=2.5cm]{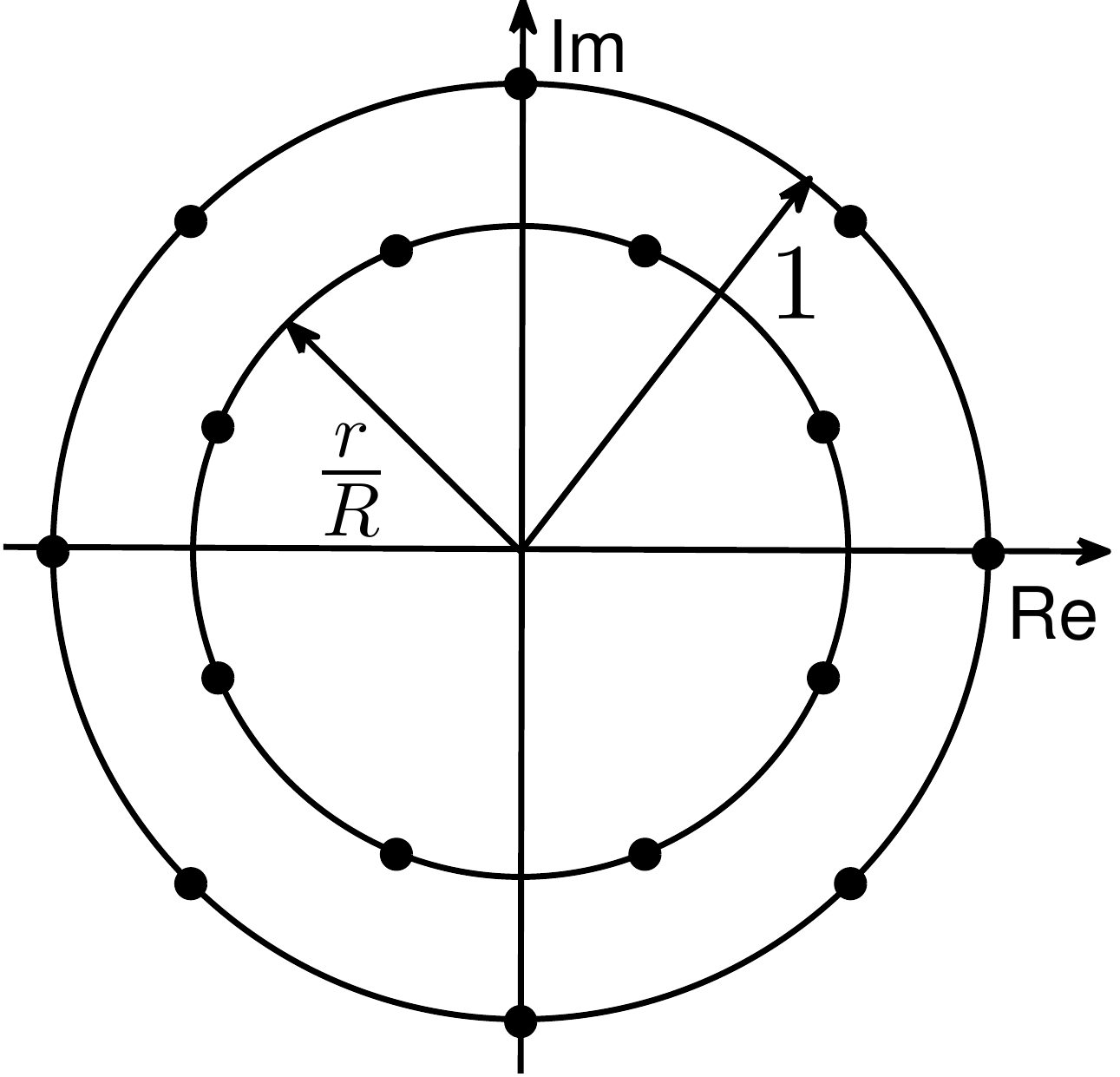}}
  \caption{Constellation diagrams of optimal $16$-ary APSK with $L=2$ for different $\frac{r}{R}$.}
  \label{APSK}
\end{figure}

To illustrate our proposed design for $N$-ary APSK constellation with $L=2$, we take the example of $N=16$ and show the optimal $\rho_2^*$ and $N_2^*$ versus $\frac{r}{R}$ in Fig. \ref{rR}. In addition, we provide in Fig. \ref{APSK} the constellation diagrams of optimal $16$-ary APSK with $L=2$, for the cases of $\frac{r}{R}=0.4, 0.5$ and $0.7$, respectively. Furthermore, the optimal set of $N$-ary APSK constellations with $L=2$ as well as their maximum MEDs (denoted by $d_{\mathrm{min},N}^*$) for the case of $N=16$ are shown in Table \ref{16APSK} at the top of next page, while those for the cases of $N=8,32$ and $64$ are provided in Tables V-VII in Appendix D. It can be observed that for each $N$, the values of $\frac{r}{R}$ are divided into several regions, each of which is associated with an optimal constellation design.\footnote{Particularly, the optimal constellation parameters as well as $d_{\mathrm{min},16}^*$ are constant over Regions 1, 3, 5 and 7. This is because for each of these regions, the optimal constellation corresponding to the left boundary value of $\frac{r}{R}$ is also feasible for the other values of $\frac{r}{R}$ in the same region.} We also show in Tables IV-VII the probability of each region for the case of i.i.d. Rayleigh fading channel with $M=2$ or $M=4$ transmit antennas, which is obtained through $10^7$ Monte Carlo trials. Moreover, for the cases of $N=2$ and $N=4$, it can be shown that for any $\frac{r}{R}\in[0,1]$, the optimal solutions to Problem (P2) correspond to BPSK (binary phase shift keying) and QPSK (quaternary phase shift keying) constellations, respectively, with $d_{\mathrm{min},2}^*=2$ and $d_{\mathrm{min},4}^*=\sqrt{2}$, respectively. For other values of $N$, the optimal constellation designs can be obtained similarly.


\begin{table*}[htb]
  \centering
  \caption{Optimal constellation set for $16$-ary APSK with $L=2$}\label{16APSK}
  \resizebox{\textwidth}{!}{
  \begin{tabular}{|c|c|c|c|c|c|c|c|}
  \hline
  Region & $\frac{r}{R}$ & $\rho_2^*$ & $N_2^*$ & $\omega_2^*$ &$d_{\mathrm{min},16}^*$ & $\mathrm{Prob}^{M=2}$& $\mathrm{Prob}^{M=4}$\\
  \hline
  1     & $[0,0.4603)$ & $0.4603$ & $5$ & $0.0182\pi$ &$0.5411$ & $0.7596$ & $0.9994$\\
  \hline
  2     & $[0.4603,0.4839)$ & $\frac{r}{R}$ & $5$ & $0.0182\pi$ &$\sqrt{\left(\frac{r}{R}\right)^2-2\left(\frac{r}{R}\right)C_{1,2}^*(N_2^*)+1}$ & $0.0245$ & $0.0002$\\
  \hline
  3     & $[0.4839,0.5176)$ & $0.5176$ & $4$ & $0.0833\pi$ &$0.5176$ & $0.0323$ & $0.0002$\\
  \hline
  4     & $[0.5176,0.5588)$ & $\frac{r}{R}$ & $4$ & $0.0833\pi$ & $\sqrt{\left(\frac{r}{R}\right)^2-2\left(\frac{r}{R}\right)C_{1,2}^*(N_2^*)+1}$ & $0.0352$ & $0.0001$\\
  \hline
  5     & $[0.5588,0.6302)$ & $0.6302$ & $8$ & $0.1250\pi$ &$0.4824$ & $0.0505$ & $0.0001$\\
  \hline
  6     & $[0.6302,0.8477)$ & $\frac{r}{R}$ & $8$ & $0.1250\pi$ &$\sqrt{\left(\frac{r}{R}\right)^2-2\left(\frac{r}{R}\right)C_{1,2}^*(N_2^*)+1}$ & $0.0844$ & $0.0000$\\
  \hline
  7     & $[0.8477,1]$ & $1$ & $8$ & $0.1250\pi$ &$0.3902$ & $0.0135$ & $0.0000$\\
  \hline
  \end{tabular}}
\end{table*}
\section{Suboptimal Solution to Problem (P2)}
Clearly, the amount of memory required to store the set of optimal $N$-ary two-ring APSK constellations at the transmitter/receiver increases with the number of regions resulting from the optimization, each corresponding to a different constellation. Moreover, the probability of each region used in fading channel may also vary significantly from one another, e.g., with $N=16$ and i.i.d. Rayleigh fading channel, it can be observed from Table \ref{16APSK} that the probability of Region 1 is much higher than that of Region 2 for the cases with $M=2$ and $M=4$ transmit antennas, respectively. This thus motivates us to design a suboptimal constellation set for $N$-ary two-ring APSK with a smaller number of regions or constellations. From Tables IV-VII, we have the following observations:
\begin{itemize}[leftmargin=*]
\item First, for each $N$, the constellation corresponding to Region 1 has the maximum MED among all the constellations; moreover, Region 1 generally has the largest probability, and the probability increases as $N$ increases;
\item Second, for each $N$, the constellation corresponding to the last region is always feasible for the values of $\frac{r}{R}$ in all the other regions.
\end{itemize}
Therefore, we propose a suboptimal constellation set design with no more than two regions. Specifically, in the suboptimal design, the range of $\frac{r}{R}$ and the constellation corresponding to the first region are identical to those in the optimal set; while the second region (if any) covers the remaining range of $\frac{r}{R}$ and uses the constellation for the last region in the optimal set.\footnote{For cases such as $N=2$ and $N=4$ where the optimal set has a single region, the suboptimal set is identical to the optimal set.} As an example, in the proposed suboptimal design for $N=16$, the constellation for $\frac{r}{R}\in[0,0.4603)$ is the same as that for Region 1 in Table \ref{16APSK}; while the remaining range of $\frac{r}{R}\in[0.4603,1]$ uses the constellation for Region 7 in Table \ref{16APSK}.

\section{Numerical Results}\label{secnum}
In this section, we provide numerical results. Consider a single-user MISO channel modeled by ${\bf{h}}^T=\sqrt{\beta}\tilde{{\bf{h}}}^T$, where $\tilde{{\bf{h}}}^T$ represents Rayleigh fading channel coefficients with i.i.d. elements $\tilde{h}_i\sim \mathcal{CN}(0,1)$, and $\beta=-90$dB denotes the channel power attenuation due to path loss.\footnote{We consider a path loss model given by $\beta=K\left(\frac{d}{d_0}\right)^{-\alpha}$ \cite{wireless}, where we assume $K=10^{-3}$, $\alpha=3$, $d_0=1$m and $d=100$m.} The average receiver noise power is set to be $\sigma^2=-94$dBm.\footnote{We assume the bandwidth of the transmitted signal is $10$MHz, the power spectral density of the receiver noise is $-174$dBm/Hz, and the noise figure due to receiver processing is $10$dB.} The average signal-to-noise ratio (SNR) is thus defined as $\mathrm{SNR}=\frac{P\beta}{\sigma^2}$. All the results below are averaged over $10^7$ independent channel realizations.
\subsection{Fixed-Rate Adaptive Receiver Constellation}\label{secnumfixed}
In this subsection, we evaluate the performance of CE precoding with fixed-rate adaptive receiver constellation, based on the optimal set of two-ring APSK constellations or the suboptimal alternative. Moreover, we examine the effect of imperfect CSIT on the performance of the proposed schemes.
\subsubsection{Performance of Optimal Constellation Set for $16$-ary Two-Ring APSK}
\begin{figure}[htb]
  \centering
  \subfigure[$M=2$]{
    \label{fig:subfig:a}
    \includegraphics[width=7.5cm]{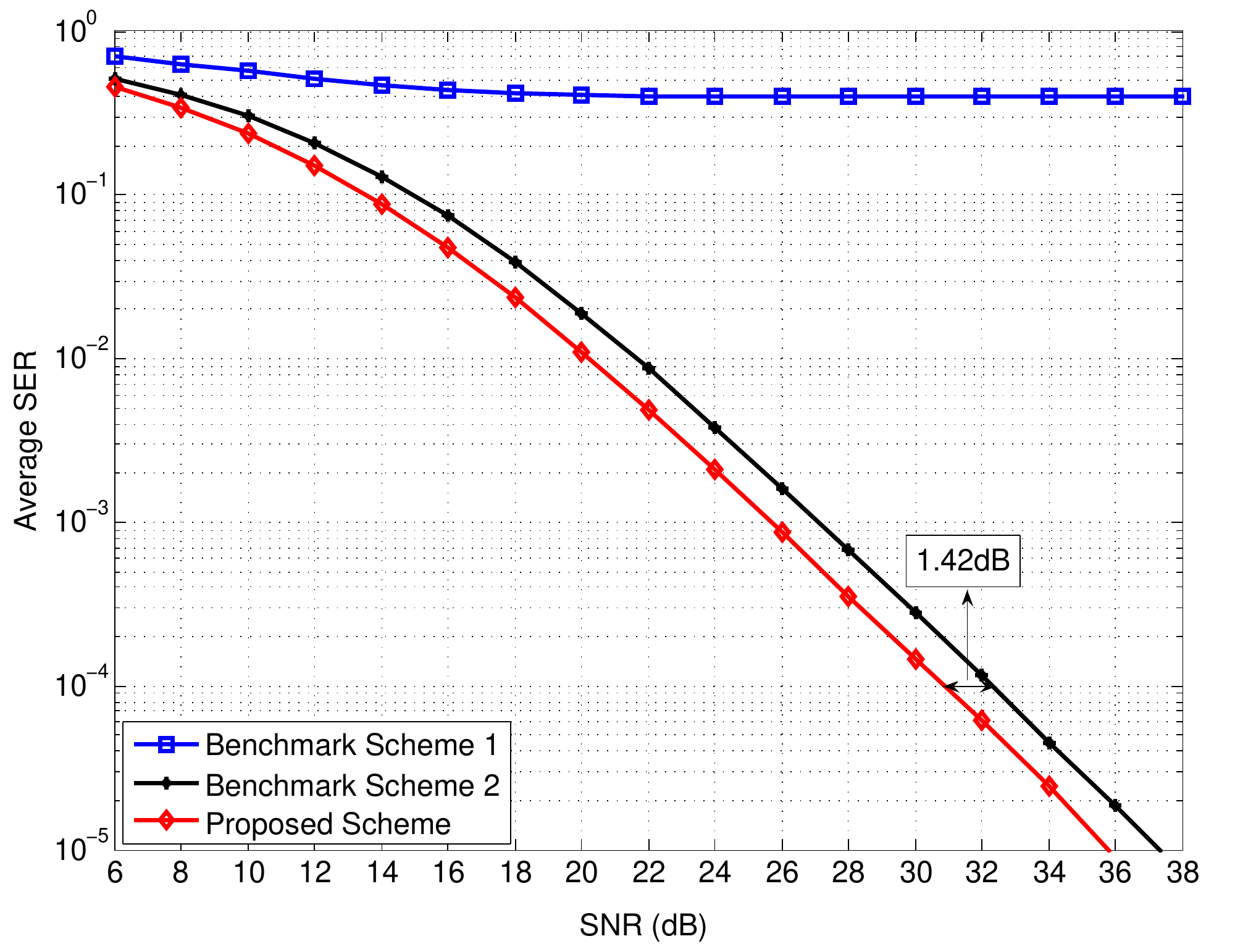}}
  \hspace{0in}
  \subfigure[$M=4$]{
    \label{fig:subfig:b}
    \includegraphics[width=7.5cm]{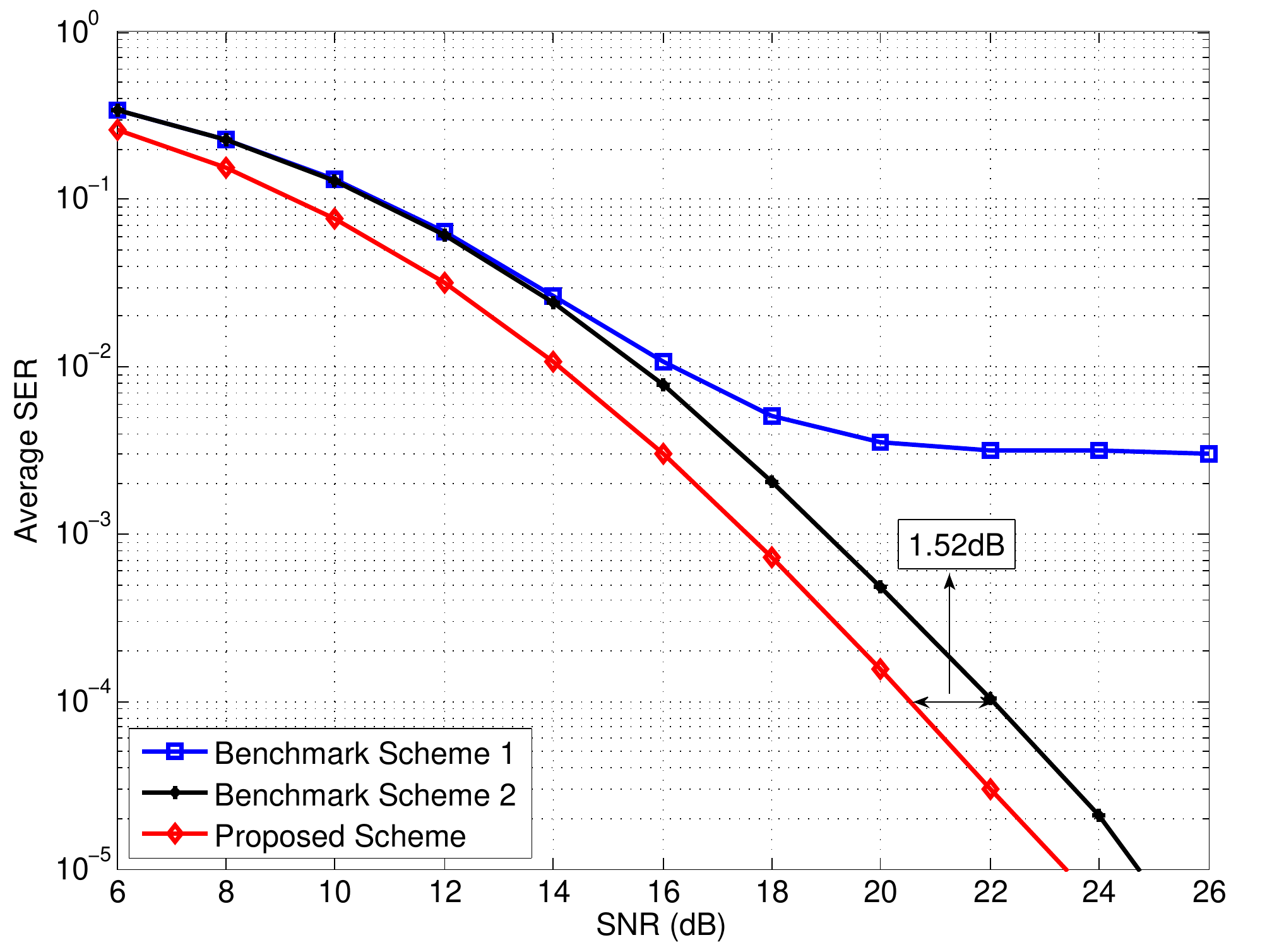}}
  \caption{Average SER comparison of fixed-rate CE transmission schemes.}
  \label{figSER}
\end{figure}

Consider the case of $N=16$, and we adopt the optimal constellation set for $16$-ary APSK with $L=2$ (as shown in Table \ref{16APSK}). In addition, we consider two benchmark schemes for performance comparison:
\begin{itemize}[leftmargin=*]
\item Benchmark Scheme 1: CE precoding with fixed rectangular 16-QAM constellation;
\item Benchmark Scheme 2: CE precoding with adaptive receiver constellation, by using rectangular 16-QAM constellation when it is feasible, i.e., $\frac{r}{R}\leq\frac{1}{3}$, and 16-PSK constellation otherwise.
\end{itemize}
In Fig. \ref{figSER}, we show the average SER comparison of our proposed scheme versus the two benchmark schemes for the case of $M=2$ or $M=4$ transmit antennas, respectively.

For both setups, it is observed from Fig. \ref{figSER} that Benchmark Scheme 1 results in severe error floor due to its infeasibility when $\frac{r}{R}>\frac{1}{3}$. It is also observed that our proposed scheme outperforms Benchmark Scheme 2 by $1.42$dB and $1.52$dB in SNR at the given SER of $10^{-4}$ for the cases of $M=2$ and $M=4$, respectively.\footnote{Since rectangular $16$-QAM has larger MED than $16$-PSK ($0.4714$ versus $0.3902$), it can be further shown that our proposed scheme outperforms CE precoding with fixed $16$-PSK constellation at the receiver, which is always feasible regardless of the channel realization.} This performance gain is due to our optimization of constellation tailored for CE transmission: when $\frac{r}{R}\leq \frac{1}{3}$, the MED of our proposed optimal constellation is $0.5411$, while that of rectangular 16-QAM is $0.4714$; when $\frac{r}{R}> \frac{1}{3}$, the MED of 16-PSK is also no larger than that of our proposed optimal constellations.

\subsubsection{Performance of Suboptimal Constellation Design}
\begin{figure}[htb]
\centering
\includegraphics[width=7.5cm]{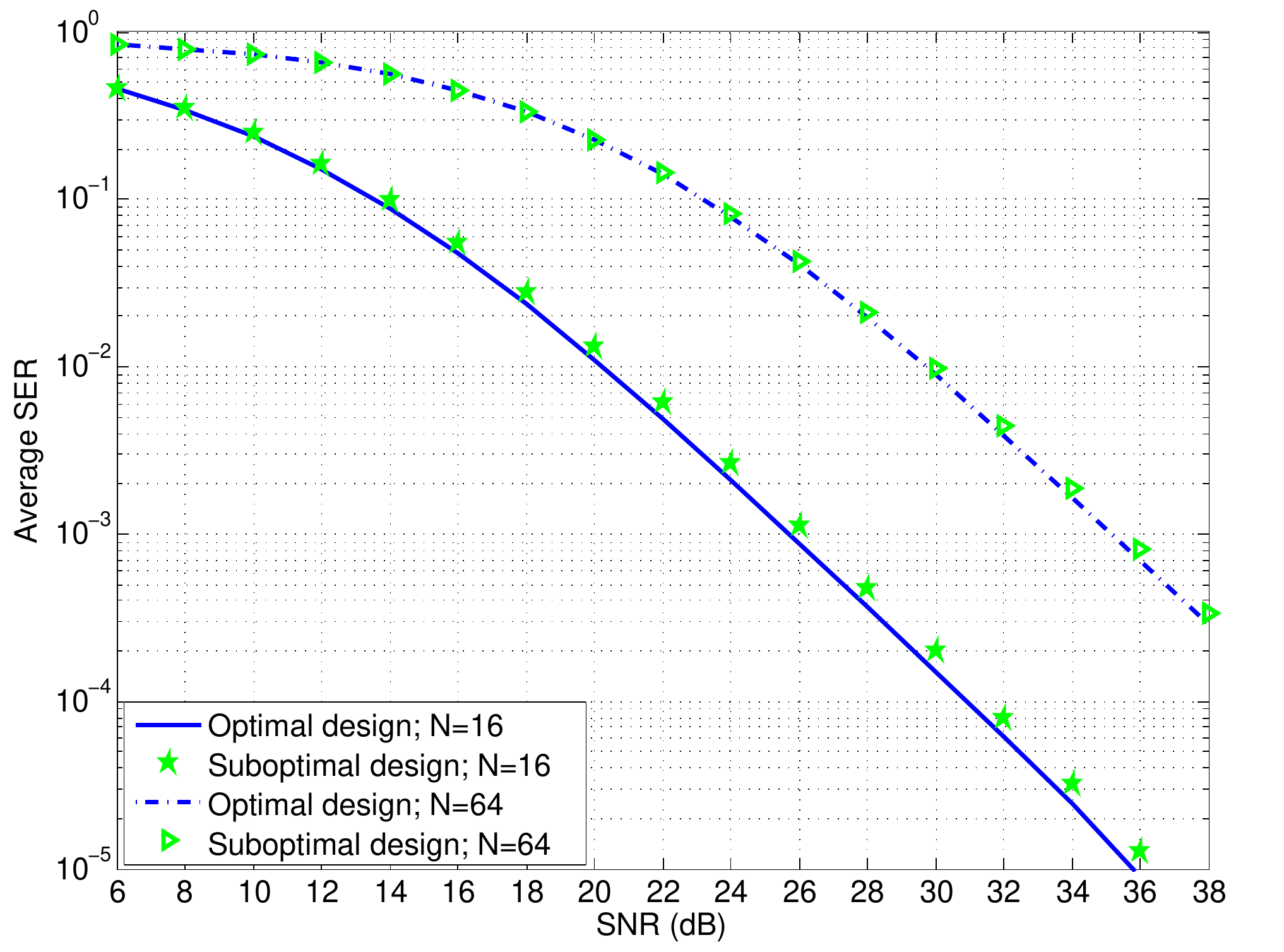}
\caption{Average SER comparison of the optimal and suboptimal constellation designs for $N$-ary APSK with $L=2$.}\label{sub2}
\end{figure}

In Fig. \ref{sub2}, we consider $M=2$ and show the average SER comparison of CE precoding with adaptive $N$-ary two-ring APSK receiver constellation based on the optimal and suboptimal constellation designs, for the case of $N=16$ or $N=64$. It is observed that for both constellation sizes, the SNR gap between the optimal and suboptimal designs at any SER is very small and almost negligible. Furthermore, this SNR gap is observed to decrease as $N$ increases, since the probability of Region 1 increases with increased $N$. The above results indicate that the proposed suboptimal constellation design of low complexity is practically efficient for fixed-rate CE transmission.
\subsubsection{Effect of Imperfect CSIT}
To realize CE precoding with the proposed fixed-rate adaptive receiver constellation, the instantaneous values of $r$ and $R$ are required to be known at the transmitter to select the constellation, which is possible since the exact channel coefficients $h_i$'s are assumed to be known at the transmitter to compute the transmitted signal phases. In this part, a time-division duplex (TDD) system is considered, where the transmitter acquires CSI via the reverse link channel training by the receiver under the assumption of  channel reciprocity.\footnote{On the other hand, applying frequency-division duplex (FDD) to CE precoding schemes requires the forward link training signal at each transmit antenna to be designed under the CE constraint. This is a challenging problem that has not been addressed yet in the literature to the best of our knowledge, thus is left as a possible direction for future work.} In practice, perfect channel estimation is generally unavailable at the transmitter, which motivates us to examine the effect of imperfect CSIT on the performance of our proposed schemes.

\begin{figure}[h]
\centering
\includegraphics[width=7.5cm]{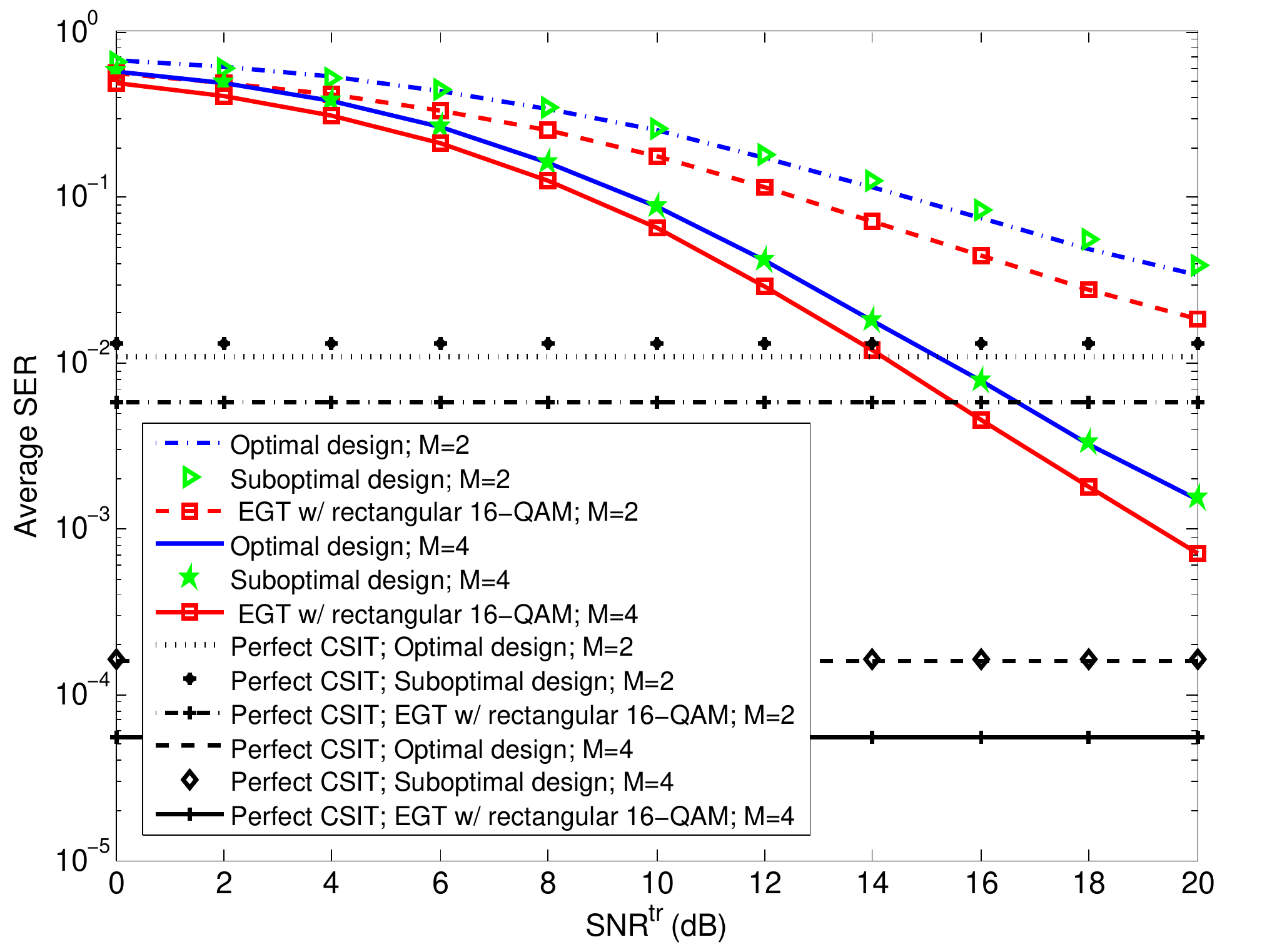}
\caption{Average SER versus average channel training SNR.}\label{imperfectCSI}
\end{figure}

Let $\hat{\mathbf{h}}$ denote the minimum mean-square error (MMSE) estimate \cite{MMSE} of the channel vector ${\bf{h}}$ at the transmitter. $\hat{\mathbf{h}}$ is modeled by $\hat{\mathbf{h}}={\bf{h}}-\Delta{\bf{h}}$, where $\Delta{\bf{h}}$ denotes the estimation error. Let $\mathrm{SNR}^{\mathrm{tr}}$ denote the average SNR of the training signal received at the transmitter in the presence of i.i.d. CSCG noise. The distribution of each element in $\Delta{\bf{h}}$ can be shown to be given by $\Delta{h_i}\sim \mathcal{CN}(0,\frac{\beta}{1+\mathrm{SNR}^{\mathrm{tr}}}),\ i=1,...,M$ \cite{MMSE}, i.e., larger $\mathrm{SNR}^{\mathrm{tr}}$ leads to more accurate CSI estimation. In Fig. \ref{imperfectCSI}, we show the average SER versus $\mathrm{SNR^{tr}}$ for our proposed CE precoding with adaptive receiver constellation based on the optimal and suboptimal designs of $16$-ary two-ring APSK for the case of $M=2$ or $M=4$ transmit antennas, respectively, where we assume $N=16$ and $\mathrm{SNR}=20$dB for the purpose of exposition. For each scheme, we also show the average SER with perfect CSIT as benchmarks. Under both setups, it is observed that the SER can be effectively improved by increasing $\mathrm{SNR^{tr}}$, which shows that the accuracy of CSIT is crucial to the performance of our proposed schemes. This is because the nonlinear CE precoding and adaptive receiver constellation selection in our proposed scheme are both based on CSIT. In addition, we show in Fig. \ref{imperfectCSI} the average SER versus $\mathrm{SNR^{tr}}$ and that with perfect CSIT for equal gain transmission (EGT) with fixed rectangular 16-QAM constellation, which is a linear non-CE precoding scheme. It is observed that the performance of our proposed schemes is almost equally sensitive to CSIT inaccuracy as that of EGT with fixed rectangular 16-QAM.

\subsection{Variable-Rate Adaptive Receiver Constellation}
In this subsection, we consider the case of variable-rate transmission with CE precoding. In general, the aim of variable-rate adaptive receiver constellation design is to maximize the spectral efficiency $\log_2N$ (in bps/Hz) by jointly adapting $N$ and $\mathcal{S}$ to the instantaneous CSI, such that the $N$-ary constellation $\mathcal{S}$ is feasible at the receiver, and the receiver SER is lower than a target value $P_e$. Similar to the fixed-rate CE transmission case, for simplicity, we replace the constraint on the exact receiver SER with that on the union bound of SER. Furthermore, for any given $\frac{r}{R}$, we assume $\mathcal{S}$ to be within the family of optimal $N$-ary two-ring APSK constellations with $N\in \mathcal{N}$, where $\mathcal{N}$ denotes the set of possible values of $N$. Therefore, we formulate the following optimization problem with given $r$, $R$, $P_e$ and $\mathcal{N}$ for the case of variable-rate CE transmission for average spectral efficiency maximization:
\begin{align}
\mbox{(P3)} &\quad \underset{N\in \mathcal{N}}{\max}\ \log_2N \nonumber \\
 \mathrm{s.t.} &\quad (N-1)Q\left(\sqrt{\frac{(Rd_{\mathrm{min},N}^*)^2}{2\sigma^2}}\right)\leq P_e,\label{P3c}
\end{align}
where $d_{\mathrm{min},N}^*$ can be readily obtained by solving Problem (P2) with Algorithm 3 for the given $\frac{r}{R}$. The optimal solution to Problem (P3) can be obtained via one-dimensional search over $N$ in a decreasing order, namely, from $N=\underset{N\in \mathcal{N}}{\max}N$ to $N=\underset{N\in \mathcal{N}}{\min}N$, until (\ref{P3c}) is satisfied.

We consider the case of $\mathcal{N}=\{2,4,8,16,32,64\}$ and $P_e=10^{-3}$ to evaluate the performance of our proposed variable-rate CE transmission scheme. For performance comparison, we also consider the following benchmark scheme:
 \begin{itemize}[leftmargin=*]
 \item Variable-rate transmission with CE precoding based on the family of rectangular QAM constellations: the constellation size for given $r$ and $R$ is chosen as the maximum $N\in\mathcal{N}$ such that the $N$-ary rectangular QAM is feasible at the receiver and yields an SER union bound below $P_e$.\footnote{There are various constellation designs for rectangular $8$-QAM with different MEDs and thus different expressions for the SER union bound. To avoid ambiguity, we use the one shown in Figure 4.3-7 (c) in \cite{digicom}.}
 \end{itemize}

\begin{figure}[htb]
  \centering
  \subfigure[Proposed scheme with two-ring APSK constellation]{
    \label{fig:subfig:a}
    \includegraphics[width=7.5cm]{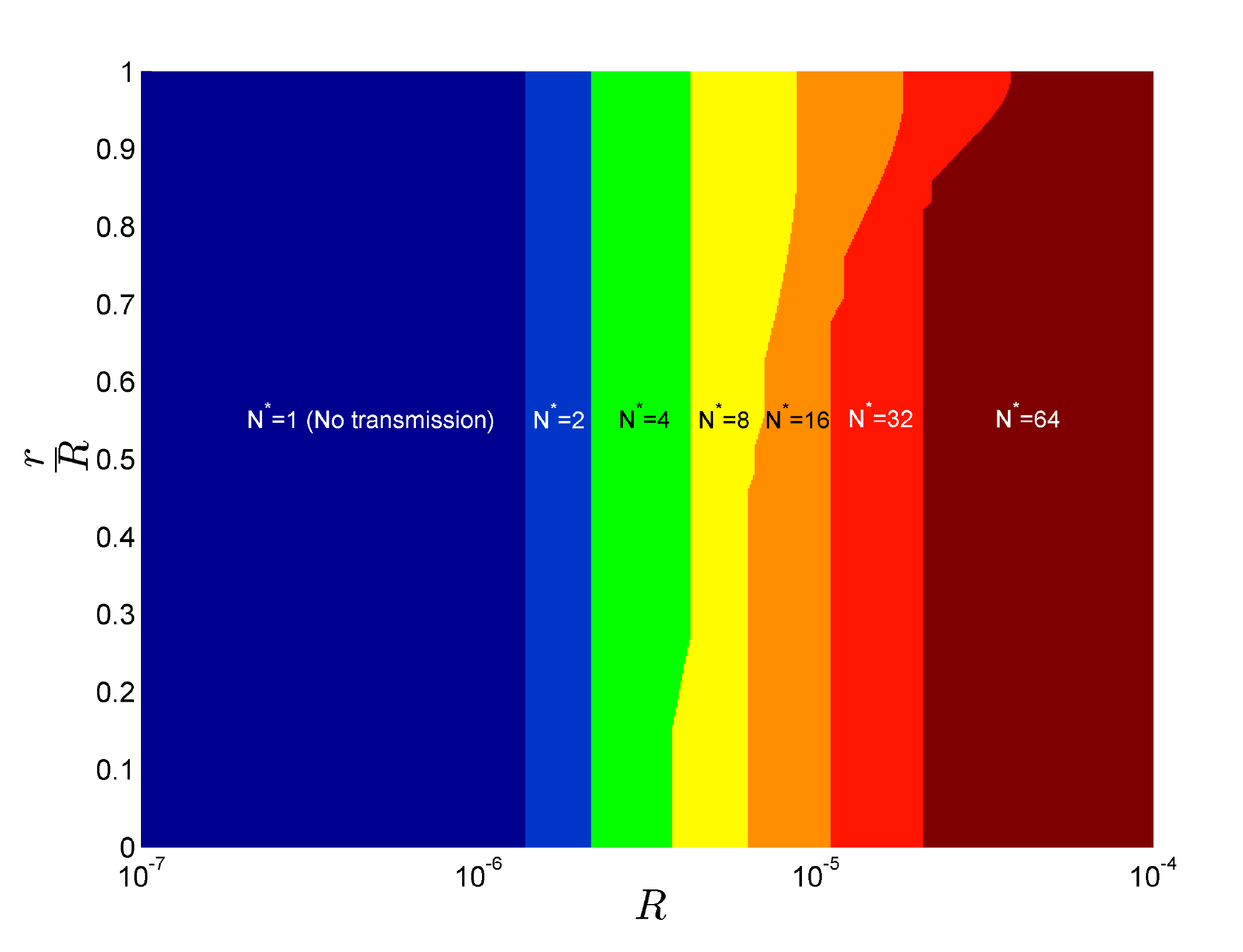}}
  \hspace{0in}
  \subfigure[Benchmark scheme with rectangular QAM constellation]{
    \label{fig:subfig:b}
    \includegraphics[width=7.5cm]{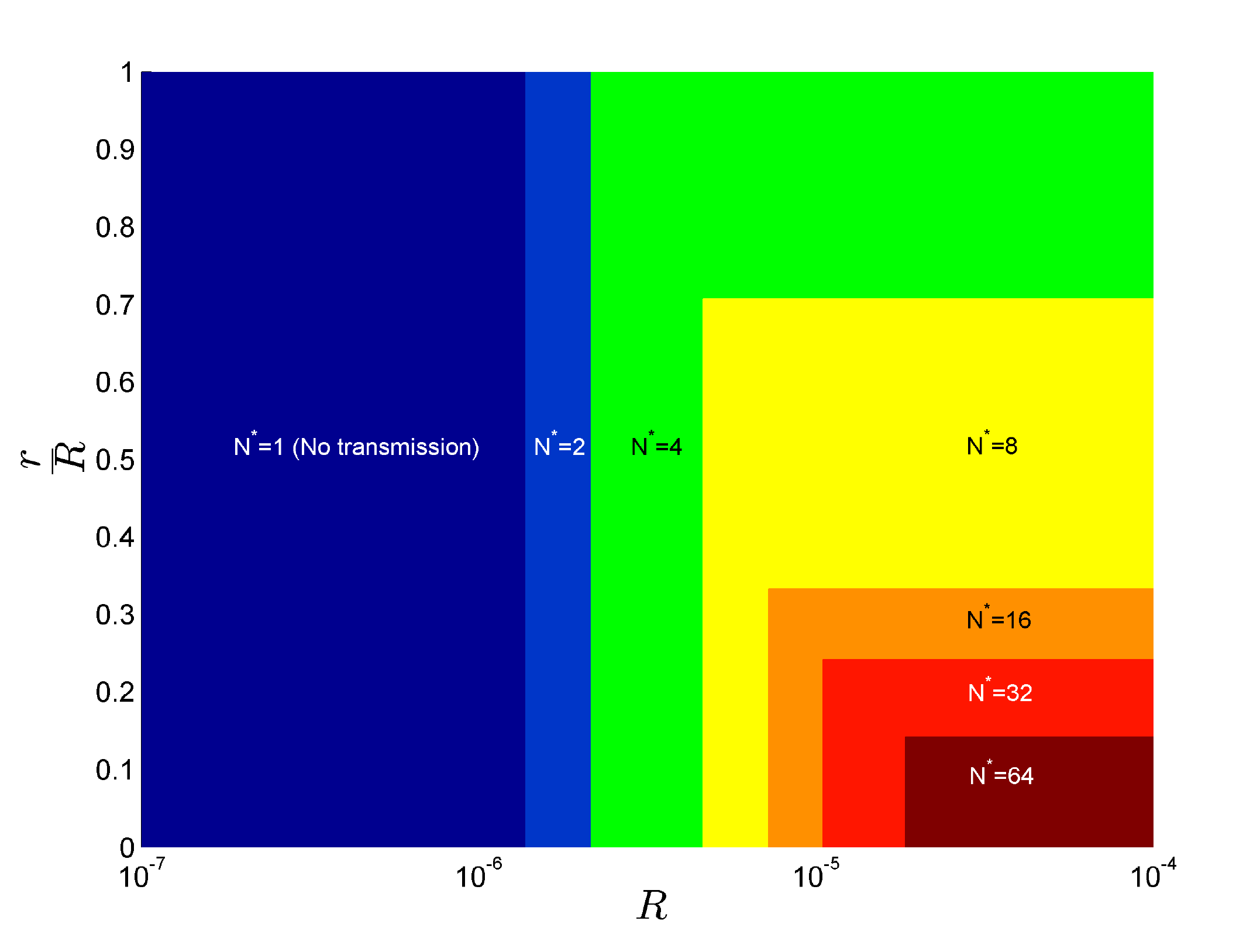}}
  \caption{Optimal constellation size $N^*$ versus $\frac{r}{R}$ and $R$.}
  \label{optimalsize}
\end{figure}

\begin{figure}[!htb]
  \centering
  \subfigure[$M=2$]{
    \label{fig:subfig:a}
    \includegraphics[width=7.5cm]{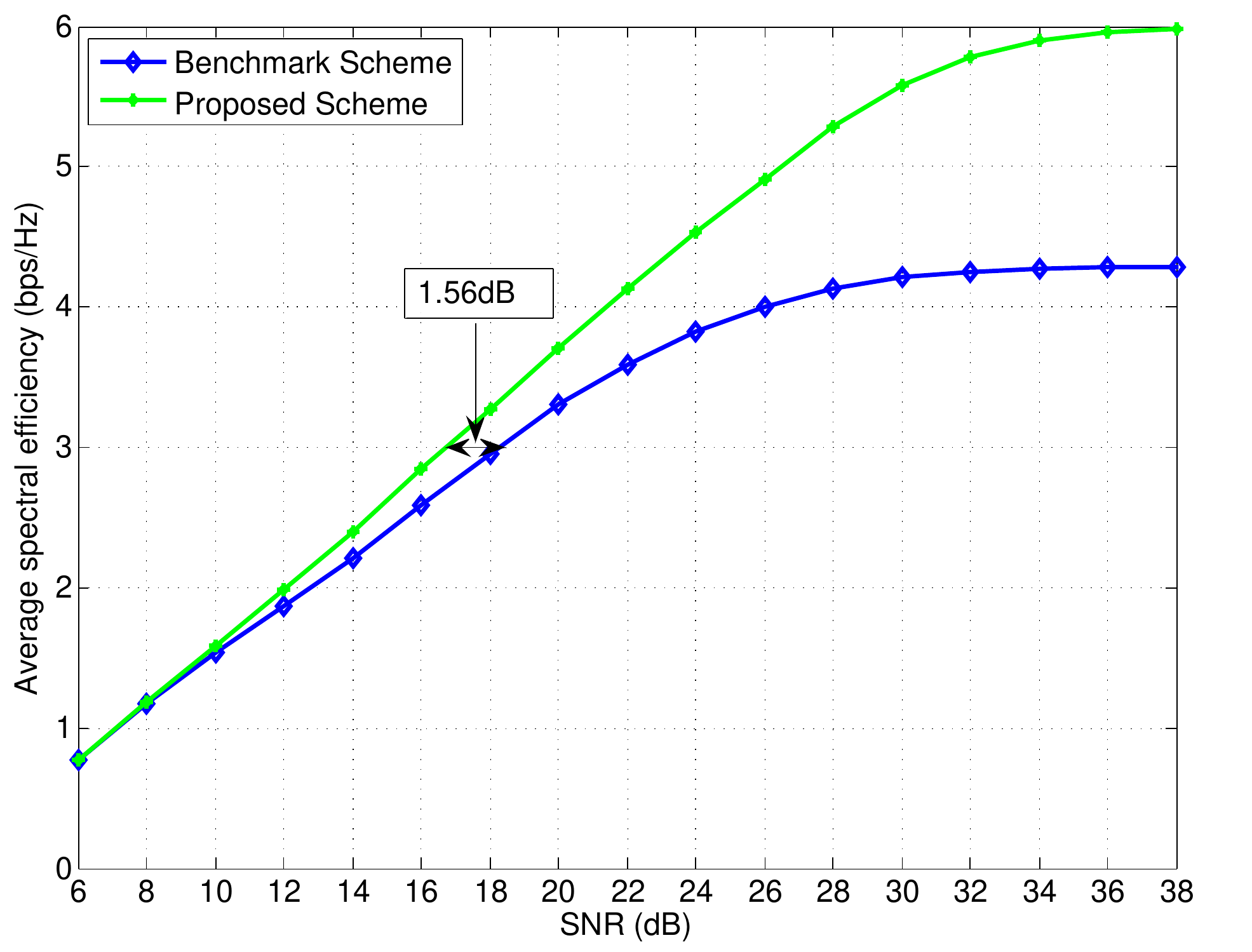}}
  \hspace{0in}
  \subfigure[$M=4$]{
    \label{fig:subfig:b}
    \includegraphics[width=7.5cm]{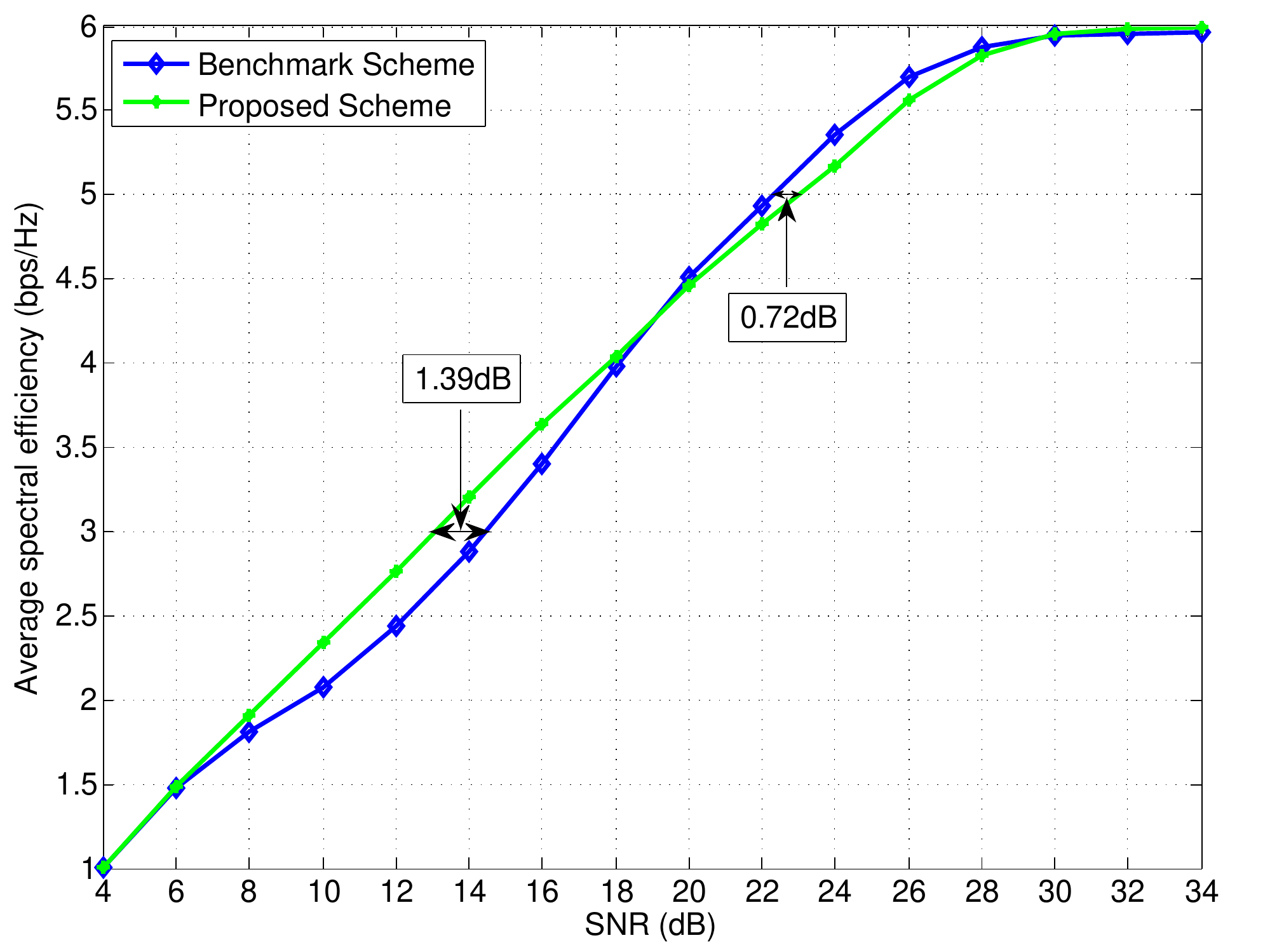}}
  \caption{Average spectral efficiency comparison of variable-rate CE transmission schemes.}
  \label{figSE}
\end{figure}

For illustration, the optimal constellation size $N^*$ versus $\frac{r}{R}$ and $R$ for both the proposed and benchmark schemes are provided in Fig. \ref{optimalsize}. Note that we let $N^*=1$ (i.e., $\log_2N^*=0$ in the objective function of (P3)) denote the case of no transmission, i.e., the SER constraint in (\ref{P3c}) cannot be met for any $N\in \mathcal{N}$. In Fig. \ref{figSE}, we compare the average spectral efficiency of our proposed scheme versus the benchmark scheme, for the case of $M=2$ or $M=4$, respectively.

For the case of $M=2$, it is observed from Fig. \ref{figSE} (a) that our proposed scheme outperforms the benchmark scheme by $1.56$dB in SNR at the average spectral efficiency of $3$bps/Hz. It is also observed that the SNR gap between the two schemes increases as the desired average spectral efficiency grows. By examining the feasibilities and MEDs of rectangular QAM constellations, this performance gain can be explained as follows. First, for $N\in\{2,4,8,16\}$, the MED of the $N$-ary rectangular QAM constellation can be shown to be no larger than that of the optimal $N$-ary two-ring APSK constellation. Furthermore, for $N\in\{8,16,32,64\}$, the probability of the $N$-ary rectangular QAM constellation being infeasible can be shown to be nonzero and increases as $N$ grows.

For the case of $M=4$, it is observed from Fig. \ref{figSE} (b) that the SNR gain of our proposed scheme over the benchmark scheme is reduced to $1.39$dB at the average spectral efficiency of $3$bps/Hz. Furthermore, the benchmark scheme even outperforms our proposed scheme by $0.72$dB of SNR at the average spectral efficiency of $5$bps/Hz. This is due to the following reasons. First, for $N\in\{8,16,32,64\}$, the probability of the $N$-ary rectangular QAM being infeasible is much lower for the case of $M=4$ than that for the case of $M=2$, e.g., rectangular $16$-QAM is infeasible for $0.3\%$ of channel realizations when $M=4$ instead of $40\%$ when $M=2$ under the assumed i.i.d. Rayleigh fading channels. Second, when rectangular $32$-QAM and $64$-QAM constellations are feasible, their MEDs ($0.3430$ and $0.2020$, respectively) are larger than those of the optimal $32$-ary and $64$-ary two-ring APSK constellations ($0.3238$ and $0.1778$, respectively). Note that the smaller MEDs of our proposed APSK constellations are due to the limited number of concentric rings (i.e., $L=2$), which can be improved by finding the solution to Problem (P1) with $L>2$, especially for large values of $N$. This is an interesting problem to be investigated in our future work.\footnote{It is worth noting that there are two new challenges for solving Problem (P1) with $L>2$ compared to the case with $L=2$. First, note that the terms $d_{l,h}$'s with $|l-h|\neq 1$ denote the inter-ring MEDs between any two signal points that are located on two non-adjacent rings, which will be present in the case of $L>2$. Additional effort is thus needed to deal with the constraint in (7) for the case of $L>2$. Second, it can be shown that the feasible set of $\{N_l\}_{l=2}^L$ in the case of $L>2$ has cardinality of $\binom{N-1}{L-1}$. Therefore, finding the optimal $\{N_l\}_{l=2}^L$ with affordable complexity is also a difficult problem for large values of $N$ and $L$.}
\section{Conclusions}\label{seccon}
This paper studied the design of adaptive receiver constellation for CE precoding in a single-user MISO flat-fading channel with arbitrary number of transmit antennas. For the case of fixed-rate transmission, an efficient algorithm was proposed to obtain the optimal APSK constellation with two rings that is both feasible and of the maximum MED for any given constellation size and instantaneous CSI. It was shown that our proposed scheme significantly improves the SER performance of CE precoding with fixed receiver constellation in fading channel. In addition, a suboptimal design for fixed-rate adaptive two-ring APSK constellations was proposed, which is of low complexity and yet achieves near-optimal SER performance. Furthermore, we proposed a variable-rate CE transmission scheme based on the family of optimal fixed-rate adaptive two-ring APSK constellation sets and examined its average spectral efficiency as compared to that based on the rectangular QAM constellations.

An appealing direction of future work is to study the optimal adaptive APSK constellation design with an arbitrary number of concentric rings, which is expected to further enhance the performance of the proposed design with two rings. It is also interesting to extend the results in this paper to the more general setups with multiple users/receive antennas.

\appendices
\section{Proof of Proposition \ref{propN2}}\label{proofpropN2}
We prove Proposition \ref{propN2} by showing that for any $N_2>\frac{N}{2}$, it is true that $d_{\mathrm{min}}^*(N_2)\leq d_{\mathrm{min}}^*(N-N_2)$. Let $\omega_2^*(N_2)$ and $\rho_2^*(N_2)$ denote the solution to Problem (P2) for any given $N_2\in\{1,...,N-1\}$. Consider two sets of $N$-ary two-ring APSK constellations denoted by $\mathcal{S}$ and $\mathcal{S}'$, respectively. Assume the numbers of points on the inner ring of $\mathcal{S}$ and $\mathcal{S}'$ are given by $N_2\in\left\{\frac{N}{2}+1,...,N-1\right\}$ and $N-N_2$, respectively. Assume the phase offsets of the inner ring of $\mathcal{S}$ and $\mathcal{S}'$ are given by $\omega_2^*(N_2)$ and $2\pi-\omega_2^*(N_2)$, respectively. Assume $\mathcal{S}$ and $\mathcal{S}'$ have the same inner ring radius $\rho_2^*(N_2)$.
For $l\in\{1,2\}$, let $d_l$ and $d_l'$ denote the intra-ring MEDs of the $l$th ring for $\mathcal{S}$ and $\mathcal{S}'$, respectively. Let $d_{1,2}$ and $d_{1,2}'$ denote the inter-ring MEDs for $\mathcal{S}$ and $\mathcal{S}'$, respectively. The MEDs of $\mathcal{S}$ and $\mathcal{S}'$ are thus given by $d_{\mathrm{min}}=\min\{d_1,d_2,d_{1,2}\}$ and $d_{\mathrm{min}}'=\min\{d_1',d_2',d_{1,2}'\}$, respectively. From (\ref{in}), we find that $d_2\leq\frac{d_{2}}{\rho_2^*(N_2)}=d_{1}'$ and $d_2<\rho_2^*(N_2){d_{1}}=d_{2}'$. Moreover, it can be observed from (\ref{adjacent}) that $d_{1,2}=d_{1,2}'$. Therefore, we have $d_{\mathrm{min}}\leq d_2\leq \min\{d_1',d_2'\}$ and $d_{\mathrm{min}}\leq d_{1,2}=d_{1,2}'$, which yield $d_{\mathrm{min}}\leq d_{\mathrm{min}}'$. Note that $d_{\mathrm{min}}=d^*_{\mathrm{min}}(N_2)$ and $d_{\mathrm{min}}'\leq d^*_{\mathrm{min}}(N-N_2)$, we thus have $d_{\mathrm{min}}^*(N_2)\leq d_{\mathrm{min}}^*(N-N_2)$. This completes our proof of Proposition \ref{propN2}.
\section{Algorithm for solving Problem (P2.1)}\label{proofalgo1}
We first observe that by symmetry, the constraint of Problem (P2.1) can be modified as $0\leq\omega_2\leq \frac{2\pi}{N_1}$, which yields $\frac{2\pi n}{N_2}+\omega_2-\frac{2\pi m}{N_1}\in\left[\frac{2\pi}{N_1}-2\pi,2\pi-\frac{2\pi}{N_2}+\frac{2\pi}{N_1}\right]\subset(-2\pi,2\pi]$. Note that for $\phi\in(-2\pi,2\pi]$, $\cos\phi$ monotonically increases with $\left||\phi|-\pi\right|$. Therefore, Problem (P2.1) is equivalent to
\begin{equation}
\begin{aligned}
\underset{0\leq \omega_2 \leq \frac{2\pi}{N_1}}{\min}\ \underset{m\in \mathcal{I}_1,n\in \mathcal{I}_2}{\max}\ f(m,n,\omega_2), \label{probphase}
\end{aligned}
\end{equation}
where $\mathcal{I}_1=\{0,1,...,N_1-1\}$, $\mathcal{I}_2=\{0,1,...,N_2-1\}$ and $f(m,n,\omega_2)=\Bigg|\left|\omega_2+\left(\frac{2\pi n}{N_2}-\frac{2\pi m}{N_1}\right)\right|-\pi\Bigg|$.

For the case of $N>2$, we have the following results for $f(m,n,\omega_2)$:
\begin{equation}\label{obf}
\begin{aligned}
&f(m,n,\omega_2)=\\
&\begin{cases}
\left|\pi+\omega_2+\frac{2\pi n}{N_2}-\frac{2\pi m}{N_1}\right|\leq f(1,0,\omega_2),\frac{2\pi n}{N_2}-\frac{2\pi m}{N_1}<-\frac{2\pi}{N_1}\\
\pi-\left|\omega_2+\left(\frac{2\pi n}{N_2}-\frac{2\pi m}{N_1}\right)\right|,\qquad\quad\frac{2\pi n}{N_2}-\frac{2\pi m}{N_1}\in[\frac{-2\pi}{N_1},0]\\
\left|\omega_2+\frac{2\pi n}{N_2}-\frac{2\pi m}{N_1}-\pi\right|\leq\max\{f(1,0,\omega_2), f(0,0,\omega_2)\},\\
\qquad\qquad\qquad\qquad\qquad\qquad\qquad\qquad\ \ \frac{2\pi n}{N_2}-\frac{2\pi m}{N_1}>0.
\end{cases}
\end{aligned}
\end{equation}

\begin{figure}[h]
  \centering
  \includegraphics[width=7.5cm]{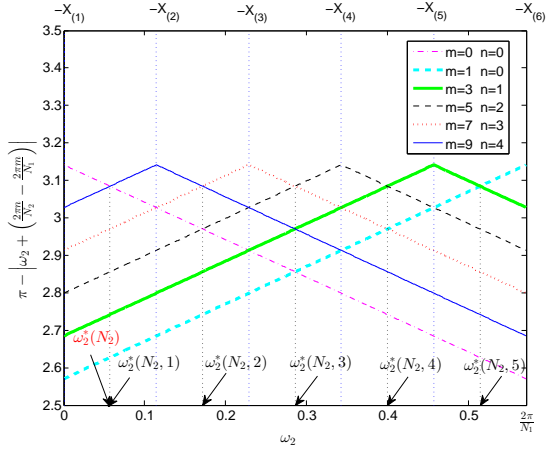}
  \caption{Illustration of Algorithm 1 with $N=16, N_2=5$.}\label{fialgo1}
\end{figure}

Based on (\ref{obf}), we further simplify Problem (\ref{probphase}) as
\begin{equation}
\underset{0\leq \omega_2 \leq \frac{2\pi}{N_1}}{\min}\ \underset{\scriptstyle m\in \mathcal{I}_1,n\in \mathcal{I}_2\atop \scriptstyle \frac{2\pi n}{N_2}-\frac{2\pi m}{N_1}\in\left[\frac{-2\pi}{N_1},0\right]}{\max}\ \pi-\left|\omega_2+\left(\frac{2\pi n}{N_2}-\frac{2\pi m}{N_1}\right)\right|.\label{probphase2}
\end{equation}
To solve Problem (\ref{probphase2}), we first select the set $\mathcal{X}=\left\{\frac{2\pi n}{N_2}-\frac{2\pi m}{N_1}\in[-\frac{2\pi}{N_1},0], m\in \mathcal{I}_1, n\in \mathcal{I}_2\right\}$ and sort the elements therein as $X_{(1)}> X_{(2)}>...> X_{(K)}$, where $K=|\mathcal{X}|$. Next, we find the locally optimal solution when $\omega_2\in\left[-X_{(k)},-X_{(k+1)}\right]$ for each $k\in\{1,2,...,K-1\}$, which is given by $\omega_2^*(N_2,k)=\frac{-X_{(k)}-X_{(k+1)}}{2}$. Finally, the globally optimal solution to Problem (\ref{probphase2}) is given by $\omega_2^*(N_2)=\frac{-X_{(k^*)}-X_{(k^*+1)}}{2}$, where $k^*=\underset{1\leq k\leq K-1}{\arg\min}\frac{X_{(k+1)}-X_{(k)}}{2}$. The corresponding optimal value to Problem (P2.1) is thus given by $C_{1,2}^*(N_2)=\cos\left(\frac{X_{(k^*)}-X_{(k^*+1)}}{2}\right)$. For the case of $N=2$, i.e., $N_1=N_2=1$, it can be easily shown that $\omega_2^*(N_2)=\frac{\pi}{N_1}$ and $C_{1,2}^*(N_2)=\cos\left(\frac{\pi}{N_1}\right)$. For consistency, we include this case in Algorithm 1 by initializing $\mathcal{X}=\left\{-\frac{2\pi}{N_1}\right\}$. An illustration of Algorithm 1 is shown in Fig. \ref{fialgo1} for the case of $N=16$ and $N_2=5$.
\section{Algorithm for finding $\rho_{2,\mathrm{I}}^*(N_2)$}\label{appendixalgo2}
\begin{figure}[!htb]
  \centering
  \subfigure[Case i, with $N_2=6$ and $\frac{r}{R}=0.25$]{
    \includegraphics[width=2.3in]{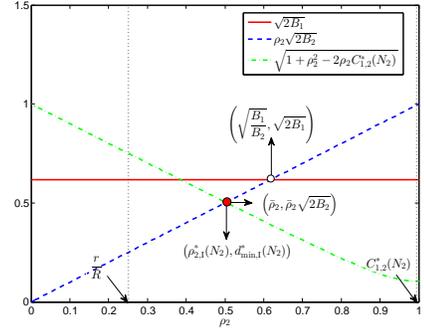}}
  \hspace{0in}
  \subfigure[Case ii, with $N_2=4$ and $\frac{r}{R}=0.25$]{
    \includegraphics[width=2.3in]{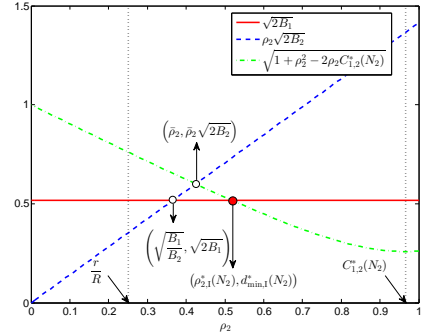}}
  \hspace{0in}
  \subfigure[Case iii, with $N_2=4$ and $\frac{r}{R}=0.55$]{
    \includegraphics[width=2.3in]{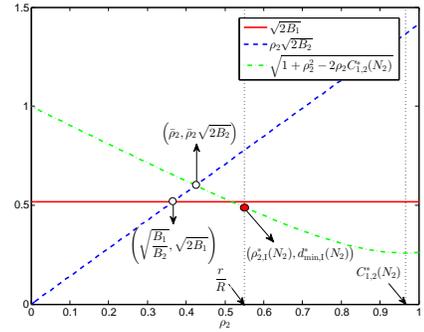}}
  \hspace{0in}
  \subfigure[Case iv, with $N_2=4$ and $\frac{r}{R}=0.45$]{
    \includegraphics[width=2.3in]{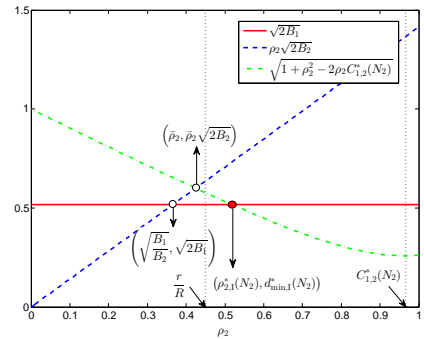}}
  \caption{Illustration of Algorithm 2 with $N=16$.}
  \label{figalgo2}
\end{figure}

We find $\rho_{2,\mathrm{I}}^*(N_2)$ by examining the objective function of Problem (P2.2), which is characterized by the intersection points of the lines given by $\sqrt{2B_1}$, $\rho_2\sqrt{2B_2}$, $\sqrt{1+\rho_2^2-2C_{1,2}^*(N_2)\rho_2}$ as well as the interval boundaries given by $\rho_2=\frac{r}{R}$ and $\rho_2=C_{1,2}^*(N_2)$. First, note that if $N_1\leq2$, we have $C_{1,2}^*(N_2)\leq0\leq\frac{r}{R}$, which contradicts the assumption of Case 2 in Section IV. Thus, we only need to consider $N_1\geq 3$ in the sequel, which yields $C_{1,2}^*(N_2)\geq \cos\left(\frac{\pi}{N_1}\right)\geq \frac{1}{2}$. Based on this result, it can be proved that when $\rho_2=C_{1,2}^*(N_2)$, $\sqrt{1+\rho_2^2-2C_{1,2}^*(N_2)\rho_2}\leq \sqrt{\frac{B_2}{2}}\leq \rho_2\sqrt{2B_2}$ and $\sqrt{1+\rho_2^2-2C_{1,2}^*(N_2)\rho_2}\leq \sqrt{\frac{B_1}{2}}\leq \sqrt{2B_1}$. Then, we denote the (left) intersection point of $\rho_2\sqrt{2B_2}$ and $\sqrt{1+\rho_2^2-2C_{1,2}^*(N_2)\rho_2}$ by $(\bar{\rho}_2,\bar{\rho}_2\sqrt{2B_2})$, where $\bar{\rho}_2$ is given by
\begin{equation}\label{brho2}
\bar{\rho}_2=\begin{cases}\frac{C_{1,2}^*(N_2)-\sqrt{C_{1,2}^{*2}(N_2)+2B_2-1}}{1-2B_2}\quad &B_2\neq\frac{1}{2}\\ \frac{1}{2C_{1,2}^*(N_2)}\ \quad &B_2=\frac{1}{2}.
\end{cases}
\end{equation}
Moreover, we find the (left) intersection point between $\sqrt{2B_1}$ and $\sqrt{1+\rho_2^2-2C_{1,2}^*(N_2)\rho_2}$ given by $\left(C_{1,2}^*(N_2)-\sqrt{C_{1,2}^{*2}(N_2)-1+2B_1}, \sqrt{2B_1}\right)$, and the intersection point between $\sqrt{2B_1}$ and $\rho_2\sqrt{2B_2}$ given by $\left(\sqrt{\frac{B_1}{B_2}},\sqrt{2B_1}\right)$. Finally, based on the above results, we discuss the solution of $\rho_{2,\mathrm{I}}^*(N_2)$ in four cases, which are summarized in Algorithm 2 shown in Table \ref{algo2}. Note that if there are multiple optimal solutions, we choose the one to maximize the SNR. For the purpose of illustration, examples of the four cases are shown in Fig. \ref{figalgo2} for the case of $N=16$.
\vspace{-2mm}
\section{Optimal Constellation Sets for $8,32$ and $64$-ary APSK with $L=2$}
\begin{table*}[!htb]
\caption{Optimal constellation set for $8$-ary APSK with $L=2$}\label{8APSK}
\centering
\resizebox{\textwidth}{!}{
\begin{tabular}{|c|c|c|c|c|c|c|c|}
\hline
Region & $\frac{r}{R}$ & $\rho_2^*$ & $N_2^*$ & $\omega_2^*$ &$d_{\mathrm{min},8}^*$ & $\mathrm{Prob}^{M=2}$ & $\mathrm{Prob}^{M=4}$\\
\hline
1     & $[0,0.1495)$ & $0.1495$ & $1$ & $0.1429\pi$ &$0.8678$&$0.2925$ & $0.9766$ \\
\hline
2     & $[0.1495,0.2705)$ & $\frac{r}{R}$ & $1$ & $0.1429\pi$ &$\sqrt{\left(\frac{r}{R}\right)^2-2C_{1,2}^*(N_2^*)\frac{r}{R}+1}$& $0.2117$ & $0.0170$\\
\hline
3     & $[0.2705,1]$ & $1$ & $4$ & $0.2500\pi$ &$0.7654$ & $0.4958$ & $0.0064$\\
\hline
\end{tabular}}
\end{table*}

\begin{table*}[!htb]
    \caption{Optimal constellation set for $32$-ary APSK with $L=2$}\label{32APSK}
    \centering
    \resizebox{\textwidth}{!}{
    \begin{tabular}{|c|c|c|c|c|c|c|c|}
    \hline
    Region & $\frac{r}{R}$ & $\rho_2^*$ & $N_2^*$ & $\omega_2^*$ &$d_{\mathrm{min},32}^*$ & $\mathrm{Prob}^{M=2}$& $\mathrm{Prob}^{M=4}$\\
    \hline
    1     & $[0,0.6764)$ & $0.6764$ & $13$ & $0.0931\pi$ &$0.3238$ & $0.9282$ & $9.9999\times 10^{-1}$\\
    \hline
    2     & $[0.6764,0.6873)$ & $\frac{r}{R}$ & $13$ & $0.0931\pi$ &$\sqrt{\left(\frac{r}{R}\right)^2-2C_{1,2}^*(N_2^*)\frac{r}{R}+1}$ & $0.0054$ & $2.5000\times 10^{-6}$\\
    \hline
    3     & $[0.6873,0.6902)$ & $0.6902$ & $12$ & $0.0167\pi$ &$0.3129$ & $0.0014$ & $0.7000\times 10^{-6}$\\
    \hline
    4     & $[0.6902,0.7074)$ & $\frac{r}{R}$ & $12$ & $0.0167\pi$ & $\sqrt{\left(\frac{r}{R}\right)^2-2C_{1,2}^*(N_2^*)\frac{r}{R}+1}$ & $0.0079$ & $3.0500\times 10^{-6}$\\
    \hline
    5     & $[0.7074,0.7583)$ & $0.7583$ & $16$ & $0.0625\pi$ &$0.2959$ & $0.0200$ & $4.2600\times 10^{-6}$\\
    \hline
    6     & $[0.7583,0.9615)$ & $\frac{r}{R}$ & $16$ & $0.0625\pi$ &$\sqrt{\left(\frac{r}{R}\right)^2-2C_{1,2}^*(N_2^*)\frac{r}{R}+1}$ & $0.0363$ & $1.8700\times 10^{-6}$\\
    \hline
    7     & $[0.9615,1]$ & $1$ & $16$ & $0.0625\pi$ & $0.1960$ & $0.0008$ &$0.0000$\\
    \hline
    \end{tabular}}
\end{table*}

\begin{table*}[!htb]
  \caption{Optimal constellation set for $64$-ary APSK with $L=2$}\label{64APSK}
  \centering
  \resizebox{\textwidth}{!}{
  \begin{tabular}{|c|c|c|c|c|c|c|c|}
  \hline
  Region & $\frac{r}{R}$ & $\rho_2^*$ & $N_2^*$ & $\omega_2^*$ &$d_{\mathrm{min},64}^*$ & $\mathrm{Prob}^{M=2}$ & $\mathrm{Prob}^{M=4}$\\
  \hline
  1     & $[0,0.8222)$ & $0.8222$ & $29$ & $0.0030\pi$ &$0.1778$ & $0.9811$ & $999.9998\times 10^{-3}$\\
  \hline
  2     & $[0.8222,0.8257)$ & $\frac{r}{R}$ & $29$ & $0.0030\pi$ &$\sqrt{\left(\frac{r}{R}\right)^2-2C_{1,2}^*(N_2^*)\frac{r}{R}+1}$ & $0.0008$ & $0.0000$\\
  \hline
  3     & $[0.8257,0.8261)$ & $0.8261$ & $28$ & $0.0119\pi$ &$0.1743$ & $0.0001$ & $0.0000$\\
  \hline
  4     & $[0.8261,0.8321)$ & $\frac{r}{R}$ & $28$ & $0.0119\pi$ & $\sqrt{\left(\frac{r}{R}\right)^2-2C_{1,2}^*(N_2^*)\frac{r}{R}+1}$ & $0.0013$& $0.0300\times 10^{-6}$\\
  \hline
  5     & $[0.8321,0.8584)$ & $0.8584$ & $32$ & $0.0312\pi$ &$0.1683$ & $0.0051$ & $0.1000\times 10^{-6}$\\
  \hline
  6     & $[0.8584,0.9903)$ & $\frac{r}{R}$ & $32$ & $0.0312\pi$ & $\sqrt{\left(\frac{r}{R}\right)^2-2C_{1,2}^*(N_2^*)\frac{r}{R}+1}$ & $0.0115$ & $0.0300\times 10^{-6}$\\
  \hline
  7     & $[0.9903,1]$ & $1$ & $32$ & $0.0312\pi$ &$0.0981$ & $0.0001$ & $0.0000$\\
  \hline
  \end{tabular}}
\end{table*}

\ \ \

\bibliographystyle{IEEEtran}
\bibliography{CEAdaptiveBib}
\end{document}